\documentclass[aps,prx,nofootinbib,twocolumn,amsmath,amssymb,superscriptaddress,notitlepage]{revtex4-2}
\usepackage{graphicx}
\usepackage{subcaption}
\usepackage{ulem}
\usepackage{enumerate}
\usepackage{bm,bbm}
\usepackage{color}
\usepackage[dvipsnames]{xcolor}
\usepackage[breaklinks,colorlinks,allcolors=blue]{hyperref}
\usepackage{textcomp}
\usepackage{multirow}
\usepackage{braket}
\usepackage{float}
\usepackage{framed}
\usepackage{adjustbox}
\usepackage{tikz}
\usepackage{mathtools}
\usepackage[all]{xy}
\usetikzlibrary{arrows.meta, positioning}
\usepackage[utf8]{inputenc}
\usepackage[T1]{fontenc} 
\usepackage{cancel}
 
\usetikzlibrary{shapes.geometric,arrows.meta,calc,math}
\newcommand{\drawgenerator}[8]{%
\xymatrix@!0{%
& #8 \ar@{-}[ld]\ar@{.}[dd] \ar@{-}[rr] & & #7 \ar@{-}[ld]  \\%
#1 \ar@{-}[rr] \ar@{-}[dd] &  & #2 \ar@{-}[dd] &            \\%
& #6 \ar@{.}[ld] &  & #5 \ar@{-}[uu] \ar@{.}[ll]       \\%
#3 \ar@{-}[rr] &  & #4 \ar@{-}[ru]                       %
}%
}

 \def\1{\mathbbm{1}}
\def\Z{\mathbb{Z}}

\newcommand{\overbar}[1]{\mkern 1.5mu\overline{\mkern-1.5mu#1\mkern-1.5mu}\mkern 1.5mu}

\makeatletter
\renewcommand*\env@matrix[1][*\c@MaxMatrixCols c]{%
  \hskip -\arraycolsep
  \let\@ifnextchar\new@ifnextchar
  \array{#1}}
  
\makeatother

\allowdisplaybreaks

\graphicspath{{figures/}}
\begin{document}

\title{Tensor Network Representations for Intrinsically Mixed-State Topological Orders
}

\author{Bader Aldossari}
\thanks{Contact author: baldossari3@gatech.edu}
\affiliation{School of Physics, Georgia Institute of Technology, Atlanta, GA 30332, USA}
\affiliation{Physics Department, King Fahd University of Petroleum and Minerals, Dhahran, Saudi Arabia}

\author{Sergey Blinov}
\thanks{Contact author: sergey.blinov@yale.edu}
\affiliation{School of Physics, Georgia Institute of Technology, Atlanta, GA 30332, USA}
\affiliation{Department of Applied Physics, Yale University, New Haven, Connecticut 06511, USA}
\affiliation{Yale Quantum Institute, Yale University, New Haven, Connecticut 06511, USA}

\author{Zhu-Xi Luo}
\thanks{Contact author: zhuxi\_luo@gatech.edu}
\affiliation{School of Physics, Georgia Institute of Technology, Atlanta, GA 30332, USA}

\date{\today}

\begin{abstract}
Tensor networks are an efficient platform for representing topological states of matter as well as computing both physical observables and information-theoretic quantities. We present a general protocol to construct tensor network representations for intrinsically mixed-state topological phases. Such phases naturally emerge in noisy long-range entangled quantum systems, and exhibit nontrivial topological phenomena without pure-state counterparts. The method exploits the isomorphism between anyon decoherence in the physical Hilbert space and anyon condensation in the corresponding Choi space. The protocol is applicable to a broad class of systems arising from decoherence of pure-state topological phases, where the incoherent noise creates excitations in the Choi space that have trivial monodromy with all excitations. Representative examples at renormalization group fixed points include $m^a e^b$ decoherence in the $\Z_N$ toric code, decohered non-Abelian quantum double of $S_3$ as well as pure $Z$/$X$ decoherence of arbitrary CSS codes. Another notable example is that of the mixed-state chiral semion, whose tensor network representation has no pure-state analogue. We then generalize the formalism beyond RG fixed points and, as an example, examine the decoherence transition in a $\mathbb{Z}_3$ toric code channel that suffers from sign problem when mapped to a statistical mechanical model. Together, these results establish tensor networks as a natural platform for analyzing noisy intrinsically mixed-state topological matter.
\end{abstract}

\maketitle

\section{Introduction \label{sec: Intro}}

Strongly correlated quantum systems can host a wide variety of exotic phases of matter, among which topologically ordered (TO) phases occupy a central place. Hallmark features of intrinsic TO include topology-dependent robust ground-state degeneracy, fractionalized excitations, and long-range entanglement \cite{zeng2019quantum}, making them both a profound theoretical discovery and a promising platform for fault-tolerant quantum computation \cite{freedman2003topological}. After decades of effort, the theoretical framework of TO is now well established \cite{zeng2019quantum,moessner2021topological}. However, realistic quantum systems are inevitably subject to decoherence—through errors, dissipation, or disorder—which can degrade these defining features and compromise their utility for encoding quantum information. Similar challenges also arise for other long-range entangled gapped systems and stabilizer codes beyond intrinsic topological phases, further underscoring the importance of extending the study of topological quantum matter beyond closed Hamiltonian systems to genuinely open quantum systems and mixed states.

Beyond immediate practical motivation, recent work has revealed that open-system settings give rise to new classes of quantum phases that cannot be realized in pure states, such as intrinsic average symmetry protected topological phases \cite{mcginley2020,deGroot2022symmetryprotected,ma2025topologicalphasesaveragesymmetries,PhysRevX.13.031016,Lee2025symmetryprotected,PhysRevB.108.155123,PRXQuantum.6.010348,guo2025newframeworkquantumphases,PhysRevB.111.L201108,guo2024locallypurifieddensityoperators} and intrinsic mixed-state topological phases  \cite{bao2023mixedstatetopologicalordererrorfield,PRXQuantum.5.020343,PRXQuantum.4.030317,PRXQuantum.6.010315,PRXQuantum.6.010313,Wang_2025,li2024replicatopologicalorderquantum,PhysRevLett.132.170602,PhysRevX.14.031044,zini2021mixedstatetqfts,zhang2025strongtoweakspontaneousbreaking1form,10.21468/SciPostPhys.17.6.167,PhysRevB.111.115142,PhysRevLett.134.070403,lessa2025higherformanomalylongrangeentanglement,li2024entanglementneededemergentanyons,kikuchi2024anyoncondensationmixedstatetopological,PhysRevLett.130.250403}. The latter include anyonic systems with transparent anyons that violate braiding non-degeneracy and hence cannot exist in pure states without conventional symmetry \cite{Kitaev_2006}.
Another class of intrinsic mixed-state examples consists of topological phases hosting a nontrivial chiral central charge. While such systems can exist in pure states, they are believed to be unrealizable as ground states of local commuting projector Hamiltonians \cite{PhysRevB.101.045137}. In the mixed-state setting, however, this obstruction need not apply \cite{Chiral_Luo,ellison2026} and a long-range correlated boundary is not necessary, so chirality must instead be diagnosed in the bulk. Together, these make the explicit representations of intrinsically mixed topological states particularly valuable, from which diagnostics such as nonlinear functionals of $\rho$ can be computed.

Tensor networks (TNs) have already proven indispensable for studying pure-state topological phases and, more broadly, area-law entangled quantum states \cite{RevModPhys.93.045003}. For many exactly solvable models in two and three dimensions, there exist TN representations \cite{Gu_2009,PhysRevB.79.085119,Schuch_2010,Bultinck_2017,PhysRevLett.122.176401,JMathPhys.54.012201,csahinouglu2021characterizing} that are fixed points under renormalization \cite{PhysRevLett.99.120601,PhysRevB.78.205116,PhysRevB.80.155131}. These exact representations  simplify the computation of correlators and have also served as powerful tools for analyzing the stability of topological phases and the nature of topological phase transitions upon deformation \cite{PhysRevB.104.235151,PhysRevB.95.235119,zhang2020non,PhysRevLett.124.130603,PhysRevLett.130.216704}.

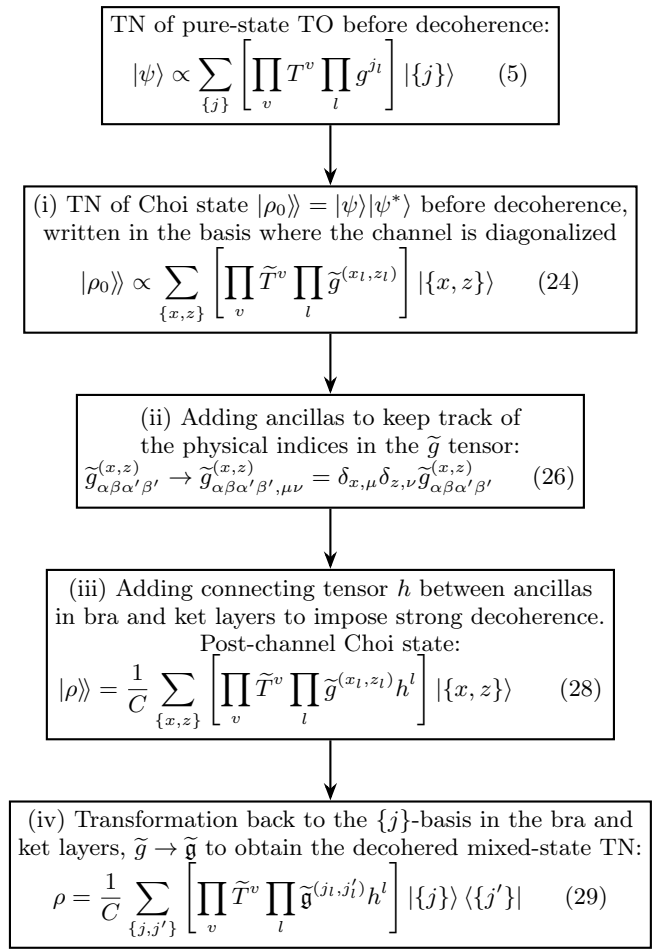
\begin{figure}
\centering
\begin{tikzpicture}[
  node distance=0.8cm and 3cm,
  align=left,
  every node/.style={draw, align=center, minimum width=1cm, minimum height=1.5cm},
  ->, >=Stealth, thick
]
\node (step0) {TN of pure-state TO before decoherence: \\ $\displaystyle \ket{\psi} \propto \sum_{\{j\}}\left[ \prod_v T^v \prod_l g^{j_l} \right]\ket{\{j\}} \text{\hspace{0.5cm}(5)}$};
\node (step1) [below=of step0] {(i) TN of Choi state  $\ket{\rho_0}\!\rangle=|\psi\rangle | \psi^*\rangle$ before decoherence, \\ written in the basis where the channel is diagonalized \\ 
$\displaystyle\ket{\rho_0}\!\rangle\propto\sum_{\{x,z\}} \left[ \prod_v\widetilde{T}^v \prod_l\widetilde{g}^{(x_l,z_l)} \right] \ket{\{x,z\}} \text{\hspace{0.5cm}(24)}$};
\node (step2) [below=of step1] {(ii) Adding ancillas to keep track of \\ the physical indices in the $\widetilde{g}$ tensor:  \\ $\displaystyle \widetilde{g}^{(x,z)}_{\alpha\beta\alpha'\beta'} \rightarrow \widetilde{g}^{(x,z)}_{\alpha\beta\alpha'\beta',\mu\nu}=\delta_{x,\mu}\delta_{z,\nu}\widetilde{g}^{(x,z)}_{\alpha\beta\alpha'\beta'} \text{\hspace{0.5cm}(26)}$};
\node (step3) [below=of step2] {(iii) Adding connecting tensor $h$ between ancillas \\ in bra and  ket layers to impose strong decoherence. \\
 Post-channel Choi state: \\
    $\displaystyle \ket{\rho}\!\rangle=\frac{1}{C}\sum_{\{x,z\}} \left[ \prod_v\widetilde{T}^v \prod_l\widetilde{g}^{(x_l,z_l)}h^l \right] \ket{\{x,z\}}$
    \hspace{0.5cm}(28)};
\node (step4) [below=of step3] {(iv) Transformation back to the $\{j\}$-basis in the bra  and \\ ket  layers, $\widetilde{g}\rightarrow \widetilde{\mathfrak{g}}$ to obtain the decohered mixed-state TN:\\ $\displaystyle \rho=\frac{1}{C}\sum_{\{ j,j' \}} \left[ \prod_v\widetilde{T}^v \prod_l\widetilde{\mathfrak{g}}^{(j_l,j'_l)}h^l \right]\ket{\{j \}}\bra{\{j'\}} \text{\hspace{0.5cm}(29)}$};
\draw (step0) -- (step1);
\draw (step1) -- (step2);
\draw (step2) -- (step3);
\draw (step3) -- (step4);
\end{tikzpicture}
\captionsetup{justification=raggedright, singlelinecheck=false}
\caption{Outline of the general TN construction procedure, demonstrated for pure-state topological phase $|\psi\rangle$ decohered in Abelian channels, i.e. the Kraus operators $K_a$ only create Abelian anyons from the topological phase. Denote the action of the each Kraus operator on the Choi-Jamiołkowski Hilbert space to be $K_a \otimes \overline{K}_a$, where $K_a$ acts on the ket and $\overline{K}_a$ acts on the bra. The Abelian channel allows for the Choi state before decoherence to be  written in the basis where $K_a \otimes \overline{K}_a$ is diagonal. The equations above are based on the $\Z_N$ examples in section \ref{sec:General Protocol}, but more general examples are available in later sections.}
\label{fig:flowchart}
\end{figure}

It is therefore natural to examine whether the TN paradigm can be extended to capture the physics of intrinsically mixed-state topological phases. While recent work has begun to address average-symmetry-protected topological phases using locally purifiable density operators \cite{xue2024tensornetworkformulationsymmetry,guo2024locallypurifieddensityoperators}, a systematic TN framework for mixed-state topological order is still lacking. 
In this work, we propose such a framework: a construction of TNs for a broad family of intrinsically mixed topological phases arising from strongly decohered or disordered pure-state topological systems. Our method exploits a duality between noise in the physical Hilbert space and anyon condensation in the Choi space, and applies whenever the decoherence operators correspond to excitations in the Choi space that have trivial self- and mutual-monodromy with respect to all excitations. This constraint is satisfied by general incoherent decoherence of Abelian topological orders, as well as a class of non-Abelian incoherent noise (see Sec.\ref{sec:S3} for an example). The resulting protocol, summarized for fixed-points in Fig. \ref{fig:flowchart}, provides a systematic path toward efficient representations of noisy topological matter.

The outline of the paper is as follows. In section \ref{sec:rev}, we provide a brief review of the quantum double formulation of $\Z_N$ topological phases on a square lattice along with standard tensor network representations of their ground states. We also describe the construction of the tensor entanglement renormalization group (TERG) coarse-graining method \cite{PhysRevLett.99.120601,PhysRevB.78.205116,PhysRevB.80.155131}, under which these tensor networks serve as fixed points.
In section \ref{sec:DS pure}, we warm up for the mixed-state case by presenting tensor network representations of pure-state topological phases that arise from bosonic anyon condensation of a parent pure-state topological phase. This formalism could be a result of separate interest complementary to the treatment discussed in \cite{PhysRevB.97.195124}, which is applicable to an overlapping but distinct set of examples.
Next, section \ref{sec:1+Av} goes over the fixed-point tensor network formulation of the first and simplest example of mixed-state topological phase: maximal decoherence of the $m$ anyon in the $\Z_2$ toric code. In section \ref{sec:General Protocol}, the general procedure will be demonstrated for $\Z_N$ topological phases subject to arbitrary incoherent decoherence of type $m^ae^b$. 
Then in section \ref{sec:extensions} we extend the discussion to the interesting examples of the decohered non-Abelian $D(S_3)$ topological phase \ref{sec:S3}, pure $Z$ or pure $X$ decoherence of generic CSS codes \ref{sec:CSS}, as well as the explicit example of the three-dimensional toric code \ref{sec:3D}, and the chiral semion case (subsection \ref{sec:chiral}). In section \ref{sec:Z3 numerics}, we extend the formalism beyond fixed-points and give an example of the usage case of this formalism by analyzing the transition point of the $\mathbb{Z}_3$ toric code under a flux-decoherence channel that faces a Monte Carlo sign problem when directly mapped to a statistical mechanical model. Finally, section \ref{sec:discussion} collects the remaining subtleties and outlook, including more discussions on the scope of the method, implications of the chiral construction and bulk-boundary relations in mixed-state topological order.

\section{Review of fixed-point tensor network representation for \texorpdfstring{$\Z_N$}{ZN}  topological phases} 
\label{sec:rev}

\begin{figure}
\centering
\includegraphics[width=0.41\textwidth]{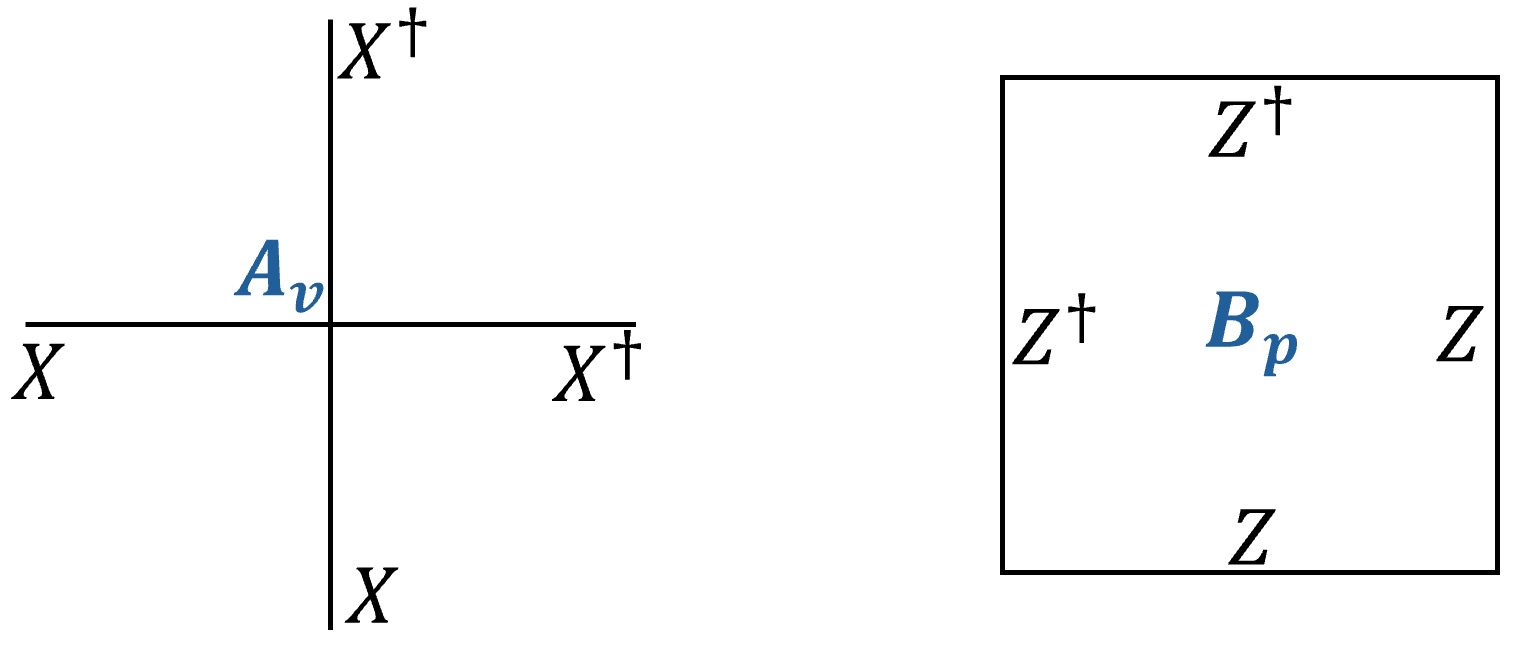}
\caption{\label{fig:AvBp} \footnotesize{Definitions of the plaquette and vertex operators}}
\end{figure}

We begin with a brief review of the $\Z_N$ topological phases \cite{Kitaev_2003} and the tensor network construction of their ground states \cite{Gu_2009,PhysRevB.79.085119}. The relevant degrees of freedom are qudits, each possessing an $N$-dimensional Hilbert space spanned by $\{\ket{j}\},$ where $j=0,\cdots, N-1$. The elementary operators acting on these qudits are the generalized Pauli operators, defined as
\begin{equation} \label{eq:general Pauli}
    X=\sum_{j=0}^{N-1} \ket{j+1}\bra{j},\;\;\;Z=\sum_{j=0}^{N-1}\omega_N^j\ket{j}\bra{j},
\end{equation}
where $\omega_N=e^{2\pi \mathrm{i} /N}$. The system resides on an $L\times L$ square lattice with periodic boundary conditions (PBC). The Hamiltonian is
\begin{equation}
    H=-\sum_v A_v-\sum_pB_p+\text{h.c.},
\end{equation}
where $A_v$ ($B_p$) is the product of the generalized Pauli $X$ ($Z$) operators on  edges nearest to vertex $v$ (plaquette $p$), with the form given in Fig. \ref{fig:AvBp}. 
All $A_v$ and $B_p$ mutually commute, leading to a ground state subspace that is stabilized by them.

One can intuitively associate an oriented string to each qudit state: A string of value $j$ with one orientation is equivalent to a string of value $N-j$ with the opposite orientation. With this perspective, the $A_v$ terms enforce a Gauss law at each vertex, requiring the net string flux through the vertex to vanish. The action of $B_p$ on such a state produces another $A_v$-stabilized state. Therefore, a ground state of $H$ is simply a superposition of all $A_v$-stabilized states that are related by $B_p$ actions. The lowest-energy excitations result from the violation of $A_v$ or $B_p$ stabilizer conditions. These violations must come in pairs, interpreted as the creation of anyons at the relevant vertexes (charges) or plaquettes (fluxes), labeled by $e$ and $m$ respectively. For example, applying $Z$ to a link produces a pair of anyons $e$ and $\bar{e}=e^{N-1}$ at the vertices living at the endpoints of said link, and likewise with pairs of flux anyons $m$ and $\bar{m}$ created at the plaquettes sandwiching the link where $X$ is applied.

\subsection{Tensor network representation of pure-state \texorpdfstring{$\mathbb{Z}_N$}{Z-N} toric code}

Tensors in a tensor network (TN) can carry two kinds of indices: (1) physical indices, which correspond to the physical degrees of freedom and take values in $\{0,1,\cdots,N-1\}$ for the $N$-qudits, and (2) virtual indices, which are to be contracted over to realize the representation of the state and are responsible for realizing entanglement through only local tensors.  The cardinality of the set a virtual index takes values in is referred to as its bond dimension, and ideally should be minimized for computational efficiency.  In this paper, physical indices will be represented as superscripts $\{i, j, k,\cdots\}$, while the virtual indices will be represented as subscripts $\{\alpha,\beta,\gamma,\cdots\}$.

\begin{figure}
    \centering
    \begin{subfigure}[b]{0.4\columnwidth}
        \centering
        \includegraphics[width=\linewidth]{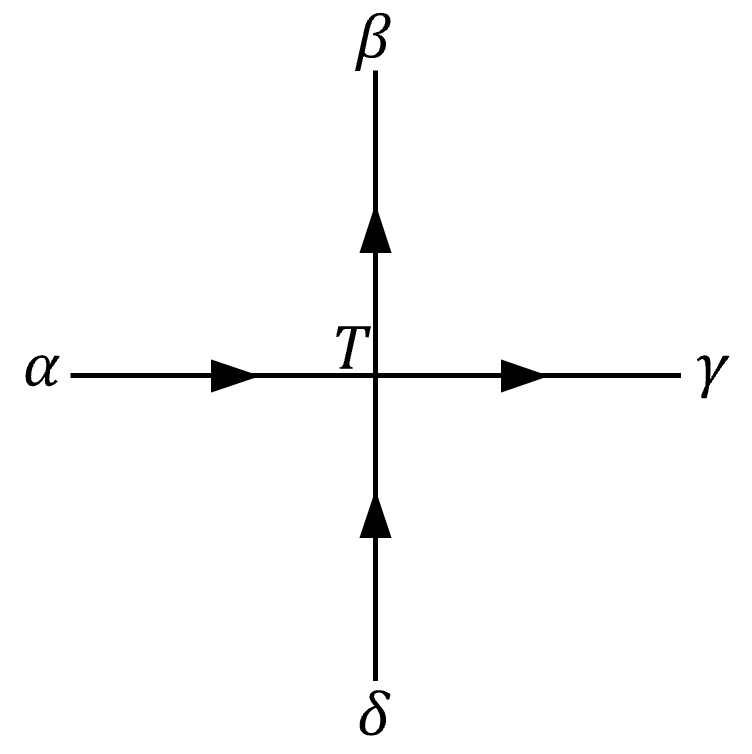}
        \caption{}
        \label{fig:Ttensor}
    \end{subfigure} \hspace{5mm}%
    \begin{subfigure}[b]{0.4\columnwidth}
        \centering
        \adjustbox{raise=1.3cm}{
        \includegraphics[width=\linewidth]{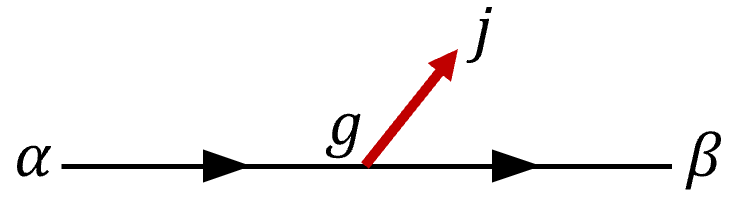}}
        \caption{}
        \label{fig:gtensor}
    \end{subfigure}
    \captionsetup{justification=raggedright, singlelinecheck=false}
    \caption{\footnotesize{(a) The $T$ tensor serves to enforce Gauss' law at each vertex $v$, with the arrows on the virtual indices indicating the string orientation convention used in the paper. (b) The $g$ tensor projects the virtual indices to the value of the physical state, as given by the index $j$ of the arrow.}}
\end{figure}

Following ref. \cite{Gu_2009}, we define a tensor $T$ at each vertex which enforces Gauss' law and contains only virtual indices with bond dimension $N$:
\begin{equation}
    T_{\alpha\beta\gamma\delta} = \begin{cases} 1 & \alpha-\beta-\gamma+\delta \; \: \text{mod }N=0  \\ 0 & \text{else} \end{cases}
\label{eq:T_pure}
\end{equation}
Here we have chosen the convention that the left and bottom legs of the tensor are pointing into the vertex, and the others away from it; see Fig. \ref{fig:Ttensor}.
We further include a tensor $g$ (Fig. \ref{fig:gtensor}) on the links that projects the neighboring virtual states onto the physical state:
\begin{equation}
    g_{\alpha\beta}^j=\delta_{\alpha,\beta,j}
\end{equation}
with $\delta_{\alpha,\beta,j}$ defined as the product of Kronecker deltas $\delta_{\alpha,\beta}\delta_{\beta,j}$. The ground state of the $\mathbb{Z}_N$ toric code model is then expressed as the sum over all physical configurations weighted by the tensor network contracted over the virtual indices (which are suppressed in the equation below):
\begin{equation}
    \ket{\psi}=\frac{1}{\sqrt{N^{M/2+1}}}\sum_{\{j\}}\left[ \prod_v T^v \prod_l g^{j_l} \right] \ket{j_1,j_2,...,j_M},
\label{eq:pure_TN}
\end{equation}
where $M=2L^2$ is the number of edges. In the notation of this paper, enclosing the TN in square brackets means the contraction of all the virtual/ancillary legs in the TN. \footnote{Other degenerate ground states orthogonal to $\ket{\psi}$ can be reached by a modification to the tensors $T$ along non-contractible loops in the lattice.}

\subsection{Tensor entanglement renormalization group procedure} \label{sec:1+Av RG}

\begin{figure}[htbp]
    \centering
    \begin{subfigure}[b]{\columnwidth}
        \centering
        \includegraphics[width=\linewidth]{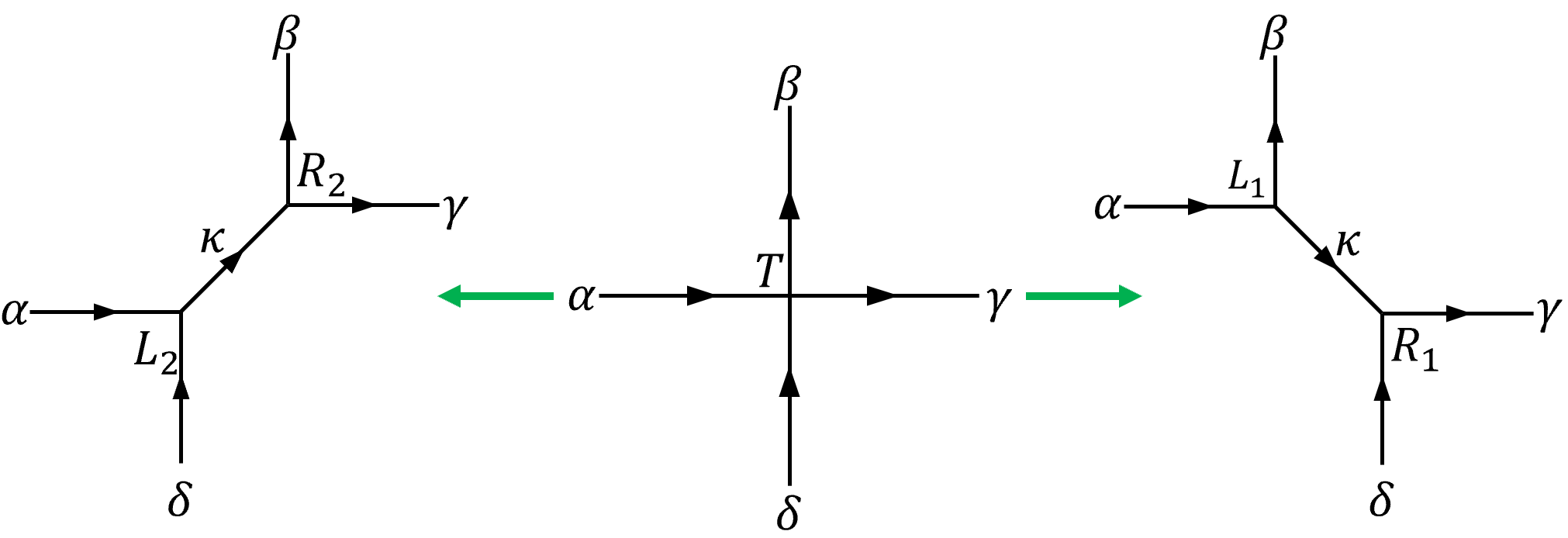}
        \caption{}
        \label{fig:Tdecomp}
    \end{subfigure}
    \begin{subfigure}[b]{\columnwidth}
        \centering
        \includegraphics[width=\linewidth]{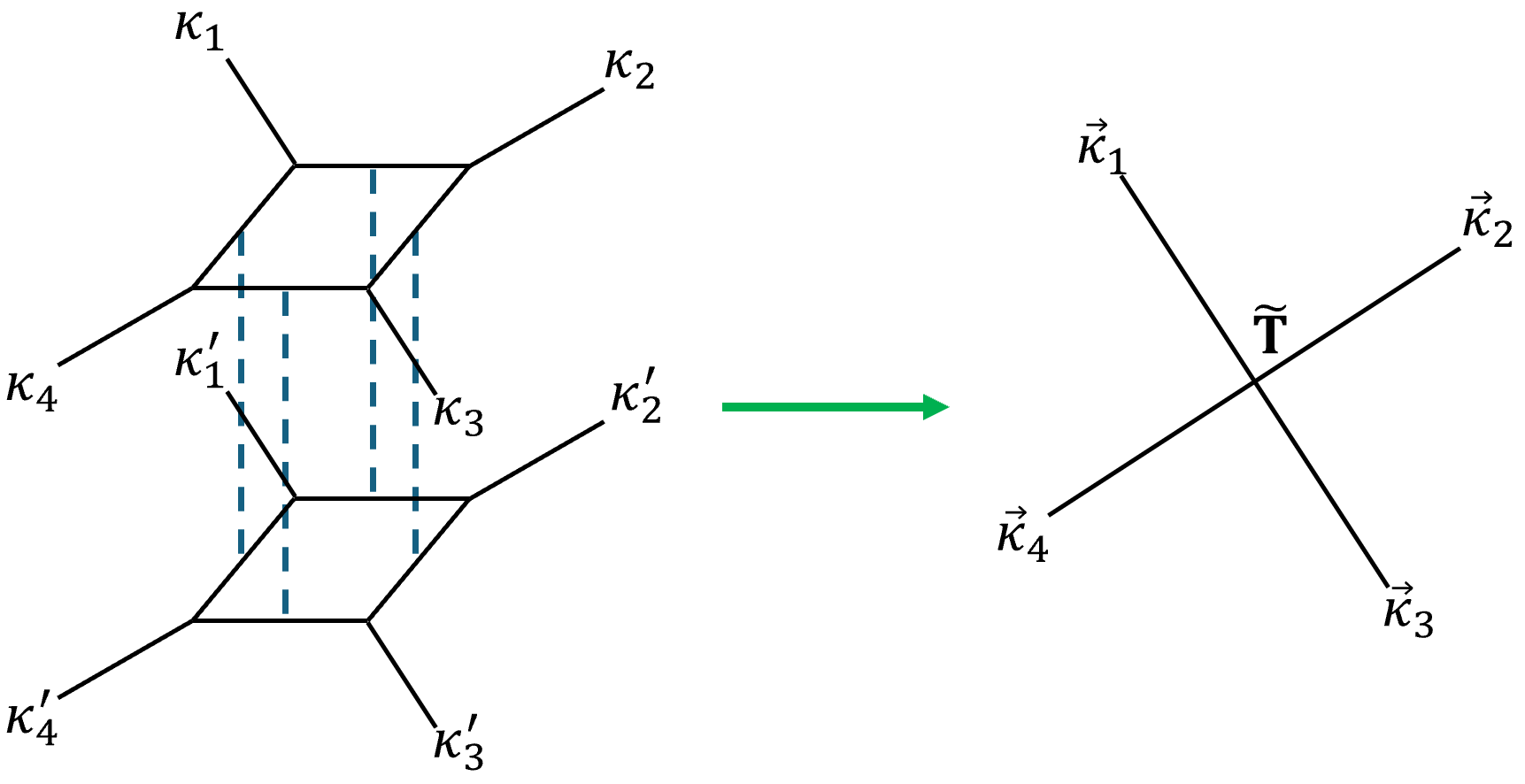}
        \caption{}
        \label{fig:squarecontraction}
    \end{subfigure}
    \begin{subfigure}[b]{\columnwidth}
        \centering
        \includegraphics[width=\linewidth]{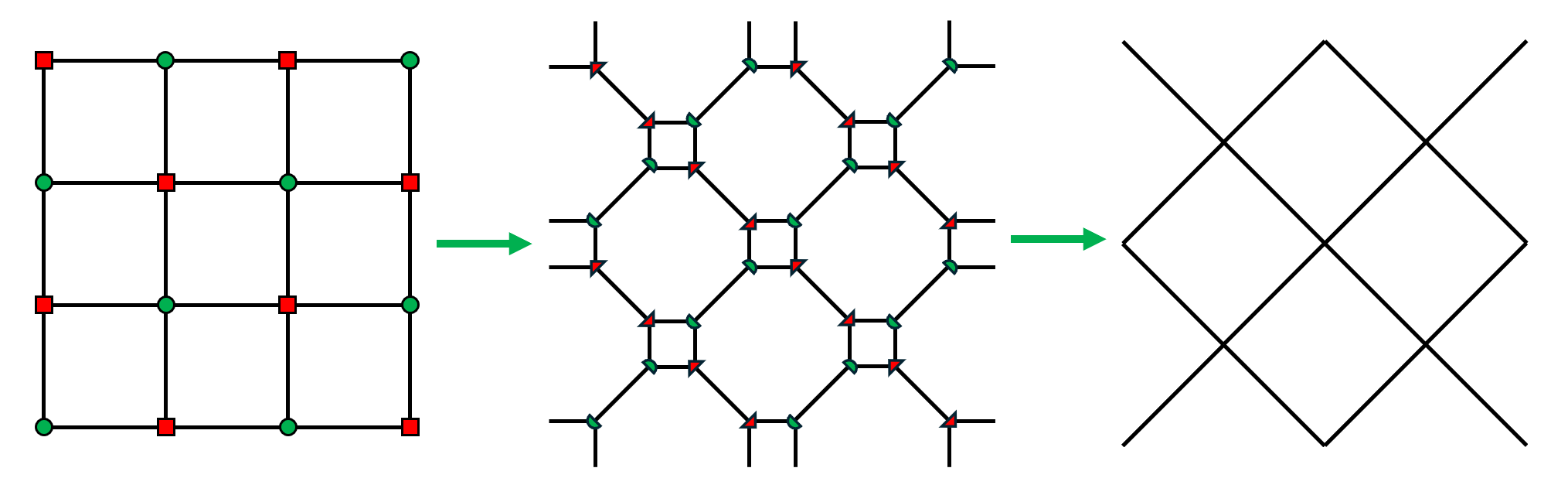}
        \caption{}
        \label{fig:RGprocedure}
    \end{subfigure}
    \captionsetup{justification=raggedright, singlelinecheck=false}
    \caption{\footnotesize{(a) Step 1 of the RG procedure. The $T$ tensor's decomposition into a product of $L$ and $R$ tensors is sublattice-dependent. (b) Step 2 of the RG procedure. The two layers are connected at the edges such that the virtual indices connected by the dashed lines must be equal. Contraction yields a new two-layer vertex tensor $\widetilde{\textbf{T}}$, which is proportional to the product of two copies of $T$. (c) Summary of the procedure of the RG on a square lattice, with final TN geometry equivalent to the original up to $\pi/4$ rotation. The red squares and green circles correspond to the two sublattices. Here, red squares (green circles) are split into $L_1$ and $R_1$ ($L_2$ and $R_2$).}}
    \label{fig:PureRG}
\end{figure}

When computing correlators $\braket{O}$ where the operator $O$ is supported in some local region(s), the cost of tensor contraction grows exponentially with the system size. Therefore, if the tensor network is invariant under renormalization 
away from the operator insertion(s), the computation efficiency can be significantly improved. The tensor network states of $\Z_N$ topological phases constructed in the previous section are fixed-points of the tensor-entanglement renormalization group \cite{PhysRevLett.99.120601,PhysRevB.78.205116} (TERG, or simply RG here) flow, illustrated in Fig. \ref{fig:PureRG}. Instead of showing that $T$ itself is invariant, we follow the standard treatment in \cite{PhysRevB.78.205116} and focus on the doubled tensor
$\tilde{T}=\sum_m T^{m}\otimes T^{m^*}$ obtained by contracting the physical index between $|\Psi\rangle$ and $\langle\Psi|$; it gives the Boltzmann weight of the
classical network whose contraction gives norms and expectation values of observables.
By comparison, the local tensor $T$ defining the wavefunction has gauge redundancies, and its singular values are not meaningful due to the fact that virtual states carry no canonical normalization. In \cite{PhysRevB.82.155138}, it was shown that  $\tilde{T}$ determines $T$ up to a unitary acting on the physical index alone: such a unitary manifestly cancels in the contraction, and it is the only remaining freedom.

Away from the support of the inserted operator $O$, the TN locally gives that of $\braket{\psi|\psi}$, achievable by contracting all the physical degrees of freedom between the bra and ket. At this point, the tensor network consists of two layers corresponding to the bra and ket,  connected by a Kronecker delta constraint at each link since $\sum_j g^j_{\alpha\beta}(g^j_{\gamma\delta})^*=\delta_{\alpha,\beta,\gamma,\delta}$. At each vertex there are two tensors from the bra and ket, which we combine as stated above into $\widetilde{T}_{\alpha\alpha'\beta\beta'\gamma\gamma'\delta\delta'}=T_{\alpha \beta\gamma\delta}T^*_{\alpha'\beta'\gamma'\delta'}$

For each (undoubled) $T$ tensor, we can split it into a product of two tensors, as shown in Fig. \ref{fig:Tdecomp}:
\begin{equation}
    T_{\alpha\beta\gamma\delta}=\sum_\kappa L_{\alpha\beta\kappa}R_{\gamma\delta\kappa}.
\label{eq:split}
\end{equation}
We choose distinct splittings for the two sublattices of the square lattice for the resultant tensor network to preserve fourfold rotation symmetry.
Viewing this splitting in equation \eqref{eq:split} as a matrix product shows that the bond dimension of $\kappa$ is at most $N^2$ for generic $L$ and $R$. A singular value decomposition is then implemented in order to drop the smallest singular values, such that the bond dimension of all legs of the tensor are equal,  $\text{dim}(\kappa)=N$. Based on the fact that $T$ enforces Gauss' law, it is straightforward to show that this procedure is exact: the dimensional reduction from the singular value decomposition only drops singular values that are equal to $0$. In particular, the following rank-3 tensors satisfy \eqref{eq:split}:
\begin{align}
    L_{1,\alpha\beta\kappa}=R_{1,\alpha\beta\kappa}=\begin{cases} 1 & \alpha-\beta \text{ mod }N=\kappa \\ 0 & \text{else} \end{cases}, \\
    L_{2,\alpha\delta\kappa}=R_{2,\alpha\delta\kappa}=\begin{cases} 1 & \alpha+\delta \text{ mod }N=\kappa \\ 0 & \text{else} \end{cases}.
\end{align}
where subscripts $1,2$ label the sublattice (Fig. \ref{fig:Tdecomp}). The index $\kappa$ (which are contracted over) can be intuitively viewed as carrying the string flux from one tensor to the other. The resultant tensor network looks like the left panel of Fig. \ref{fig:squarecontraction}.

Finally, we contract over the square in the middle to arrive at the right panel of Fig. \ref{fig:squarecontraction}. It is easy to check that the new tensor $\widetilde{\textbf{T}}$ (bold is used to indicate the post-RG tensor) for the doubled layer has exactly the same form as the original doubled tensor $\widetilde{T}$. 
We thus obtain a coarse-grained tensor network rotated by $45$ degrees and the TN in \eqref{eq:pure_TN} is an exact fixed point of the RG flow. The full procedure is summarized in Fig. \ref{fig:RGprocedure}.

\section{Example of anyon condensation in the tensor network formalism } \label{sec:DS pure}

Before moving to mixed-states, we first present the tensor network representation of pure-state topological orders that result from boson condensation in another $\Z_N$ topological phase, as this will be instructive for later discussions. Another TN representation of anyon condensation in $\Z_N$ pure-state topological phases is given in \cite{PhysRevB.95.235119,PhysRevB.97.195124}. However, the formalism is not obviously extendable to general Abelian cases such as condensation from  $\Z_N\times\Z_N$-type topological order\footnote{In \cite{PhysRevB.97.195124}, the symmetry actions on the physical and ancillary degrees of freedom of transfer matrix fixed-points determine which anyons are condensed or confined. For condensation in more complicated topological orders such as $\Z_N\times \Z_N$, there may not exist a compatible set of operators that realize the necessary symmetry actions on the ancillary indices in particular.}. The method in this section, by comparison, is expected to hold more generally - we will visit in section \ref{sec:DS mixed} an example of $e^2m^2\times e^2m^2$ condensation from $\Z_4\times \Z_4.$  

\begin{figure}
    \centering
    \includegraphics[width=0.55\columnwidth]{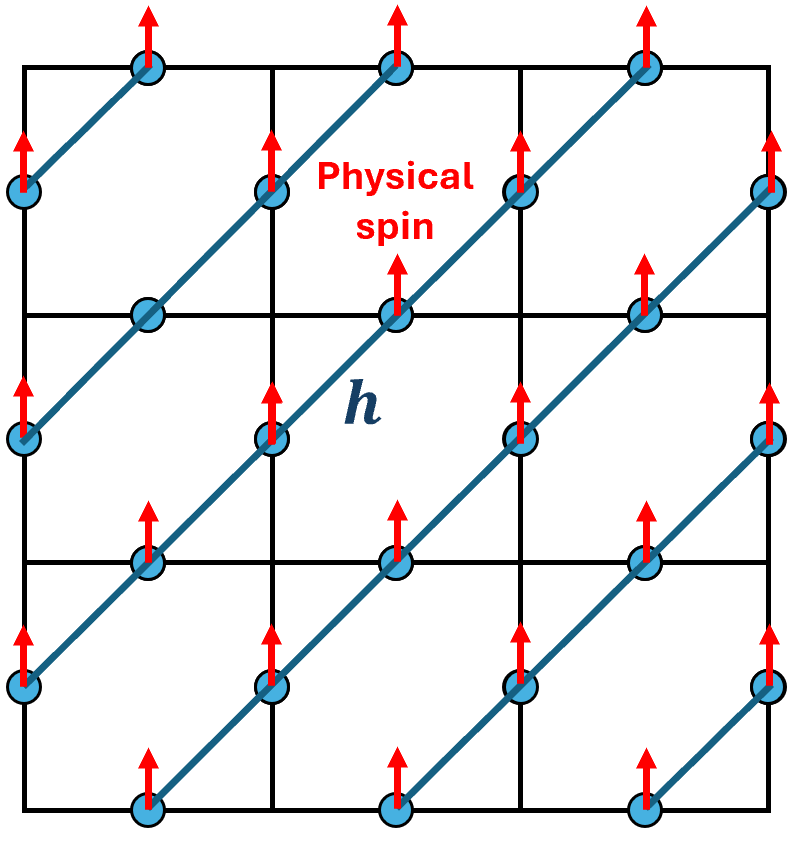}
    \captionsetup{justification=raggedright, singlelinecheck=false}
    \caption{\footnotesize{Tensor network of the double semion ground state. $h^l$ connects the virtual indices $(\mu_l, \nu_{l-a})$.}}
    \label{fig:PureDS}
\end{figure}

\begin{figure}
    \centering
    \begin{subfigure}[b]{0.45\columnwidth}
        \includegraphics[width=\linewidth]{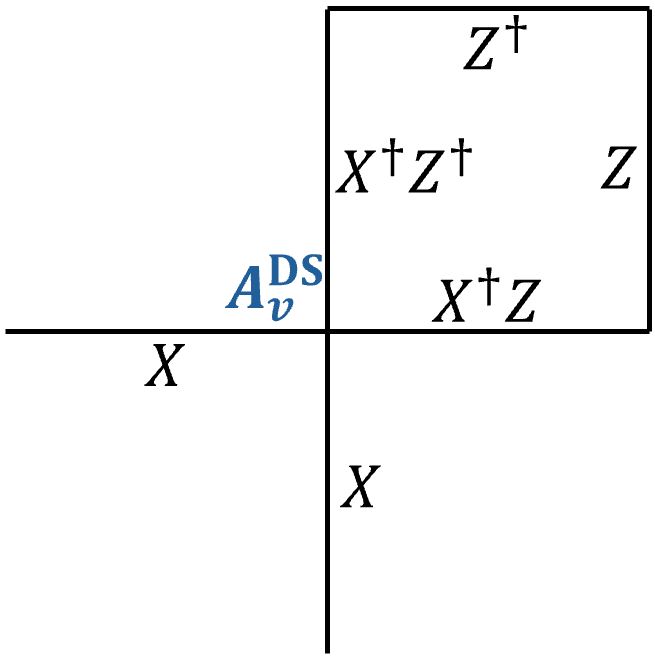}
    \caption{}
    \end{subfigure}
    \begin{subfigure}[b]{0.45\columnwidth}
        \adjustbox{raise=.9cm}{\includegraphics[width=0.55\linewidth]{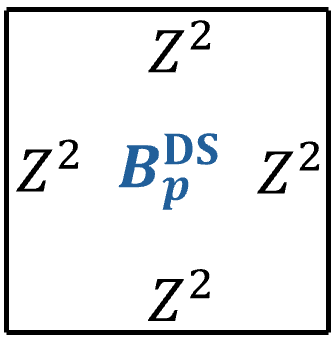}}
    \caption{}
    \end{subfigure}
    \begin{subfigure}[b]{0.55\columnwidth}
        \includegraphics[width=\linewidth]{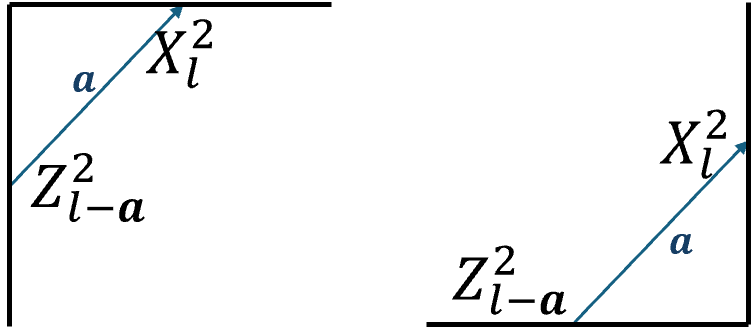}
    \caption{}
    \end{subfigure}
    \captionsetup{justification=raggedright, singlelinecheck=false}
    \caption{\footnotesize{Operators that generate the stabilizer group of the double semion ground state. Operators in panel (c) create a pair of dyons $e^2m^2$; these operators being stabilizers of the ground state is equivalent to the statement that the dyon has been identified with the vacuum.}}
    \label{fig:DS stabilizers}
\end{figure}

We start with the example of the tensor network representation of the double semion ground state obtained from the tensor network representation of $\Z_4$ topological order.
Let $N=4$ and consider the condensation of the dyon $e^2m^2$. Such self-bosonic dyons can be created by local operators in Fig. \ref{fig:DS stabilizers}(c), which mutually commute with each other. We can therefore choose the basis where these operators are diagonalized. Denote this basis as $\ket{x,z},\;x,z\in\{0,1\}$, where $x$ and $z$ label the powers of eigenvalues of $X^2$ and $Z^2$ respectively, i.e. $X^2\ket{x,z}=(-1)^x\ket{x,z}$, and similarly for $Z^2$. Let $U^{(x,z)}_j$ denote the elements of the change-of-basis matrix from $\ket{x,z}$ to $\ket{j}$, such that 
\begin{equation}
g^{(x,z)}_{\alpha\beta}\equiv\sum_j U^{(x,z)}_j g^j_{\alpha\beta}.
\end{equation}
In this basis, the TN of the $\Z_4$ ground state becomes:
\begin{equation} \label{eq:basis change pure}
    \ket{\psi}=\sum_{\{ x,z \}} \left[ \prod_v T^v \prod_l g^{(x_l,z_l)} \right] \ket{(x_1,z_1),(x_2,z_2),...}.
\end{equation}
Enforcing the condensation of $e^2m^2$ thus amounts to requiring in the tensor network:
\begin{equation}
    (x_l+z_{l-\textbf{a}})\text{ mod }2=0\;\;\forall l.
\end{equation}
To impose this condition, we first add $\Z_2$-valued ancillae $\mu, \nu$ on each link and extend the definition of $g$ to project onto their allowed eigenvalues\footnote{This step can generally be skipped if one wishes to represent the state through a contraction of physical indices, rather than explicitly summing over physical configurations of the degrees of freedom.}:
\begin{equation} \label{eq:g extend pure}
    g^{(x,z)}_{\alpha\beta}\rightarrow g^{(x,z)}_{\alpha\beta,\mu\nu}=\delta_{x,\mu} \delta_{z,\nu} g^{(x,z)}_{\alpha\beta},
\end{equation}
Note that there is no summation over $x,z$ in the equation above. We further add another tensor $h$ of the following form:
\begin{equation}
    h_{\mu_l\nu_{l-\textbf{a}}}=\begin{cases}
        1 & (\mu_l+\nu_{l-\textbf{a}})\text{ mod }2=0,\\0&(\mu_l+\nu_{l-\textbf{a}})\text{ mod }2\neq0.
    \end{cases}
\end{equation}
Upon proper normalization, we arrive at the tensor network representation of the double semion topological order shown in Fig. \ref{fig:PureDS}:
\begin{equation}
    \ket{\psi_\text{cond}}\propto\sum_{\{ x,z \}} \left[ \prod_v T^v \prod_l g^{(x_l,z_l)} h^l \right] \ket{\{x,z\}},
\label{eq:psi_cond}
\end{equation}
It can be easily verified that $\ket{\psi_\text{cond}}$ is stabilized by the appropriate stabilizers of the double semion state\footnote{Note that condensation requires the anyon to have bosonic self-statistics in order to be identified with the vacuum, and whether $e^2m^2$ is a boson or not depends on $N$.} \cite{PRXQuantum.3.010353}, which we review in Fig. \ref{fig:DS stabilizers}.

The method above can be extended to the condensation of an arbitrary anyon with bosonic self-statistics from the $\Z_N$ topological phase. The operator that creates the shortest string with said anyon at its endpoints is $X^a_lZ^b_{l-\textbf{a}}$, with $ab\text{ mod }N=0$. Condensation requires all operators in the group $S_{m^ae^b}=\braket{\{X^a_lZ^b_{l-\textbf{a}}\} }$ to be stabilizers. After repeating the steps (\ref{eq:basis change pure} - \ref{eq:g extend pure}) generalized to $\Z_N$ and $m^ae^b$, we can include the additional tensor $h$:
\begin{equation}
    h_{\mu_l\nu_{l-\textbf{a}}}=\begin{cases}
    1 & (a\mu_l+b\nu_{l-\textbf{a}})\text{ mod }N=0,\\0&(a\mu_l+b\nu_{l-\textbf{a}})\text{ mod }N\neq0,
    \end{cases}
\end{equation}
which, when included in the TN, gives the state post-condensation as in \eqref{eq:psi_cond}, but with the virtual index $\mu$ $(\nu)$ and physical index $x$ $(z)$ taking values in $\{0,1,\dots,d_a-1\}$ $(\{0,1,\dots,d_b-1\})$, where $d_a$ $(d_b)$ is the smallest positive integer such that $\left(\omega_N^a\right)^{d_a}=1$ $\left(\left(\omega_N^b\right)^{d_b}=1\right)$. The basis is diagonalized with respect to both $X^a$ and $Z^b$ (which is possible since $m^ae^b$ is a boson by assumption).

\section{\texorpdfstring{$\Z_N$}{ZN} topological order with Pauli noise} \label{sec:General Protocol}

\subsection{Flux decoherence in Toric Code} \label{sec:1+Av}

In this section, we present the simple example of tensor network representation for intrinsically mixed-state topological order obtained from the flux-channel decoherence of $\Z_2$ toric code \cite{PRXQuantum.5.020343,bao2023mixedstatetopologicalordererrorfield,PRXQuantum.4.030317,PhysRevLett.134.070403,PhysRevLett.130.250403}. Denote the initial pure state density matrix by $\rho_0$. 
Consider the maximum decoherence channel that proliferates $m$ anyons:
\begin{equation}
   \rho=\mathcal{N}\left[\rho_0 \right]=\circ_l\;\mathcal{N}_l\left[ \rho_0 \right] ,\quad \mathcal{N}_l\left[(\cdot)\right]=\frac{1}{2}\left( (\cdot)+X_l(\cdot) X_l \right),
\end{equation}
where $\circ$ denotes composition. The channel commutes with vertex operators $A_v$, leading to $\rho$ being the maximally mixed-state in the subspace where $A_v$ acts as the identity:
\begin{equation} \label{eq:1+Av}
    \rho=\frac{1}{2^{M/2+1}}\prod_v\frac{1+A_v}{2},
\end{equation}
where $M=2L^2$ is the number of edges as before, and the $1$ after the product is shorthand for the identity acting on the links connected to $v$. While this state does not encode any topological quantum information, it is still topologically non-trivial and exhibits strong-to-weak spontaneous symmetry breaking of 1-form symmetries \cite{zhang2025strongtoweakspontaneousbreaking1form}.

\begin{figure}
    \centering
    \begin{subfigure}[b]{0.45\columnwidth}
        \centering
        \includegraphics[width=\linewidth]{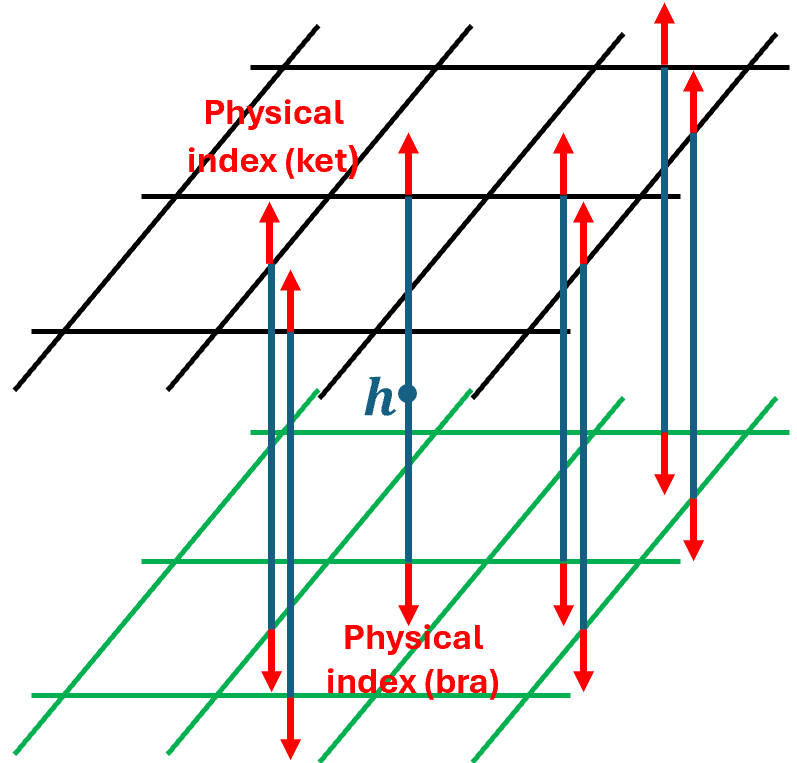}
        \caption{}
        \label{fig:1plusBP}
    \end{subfigure}
    \begin{subfigure}[b]{0.45\columnwidth}
        \centering
        \includegraphics[width=\linewidth]{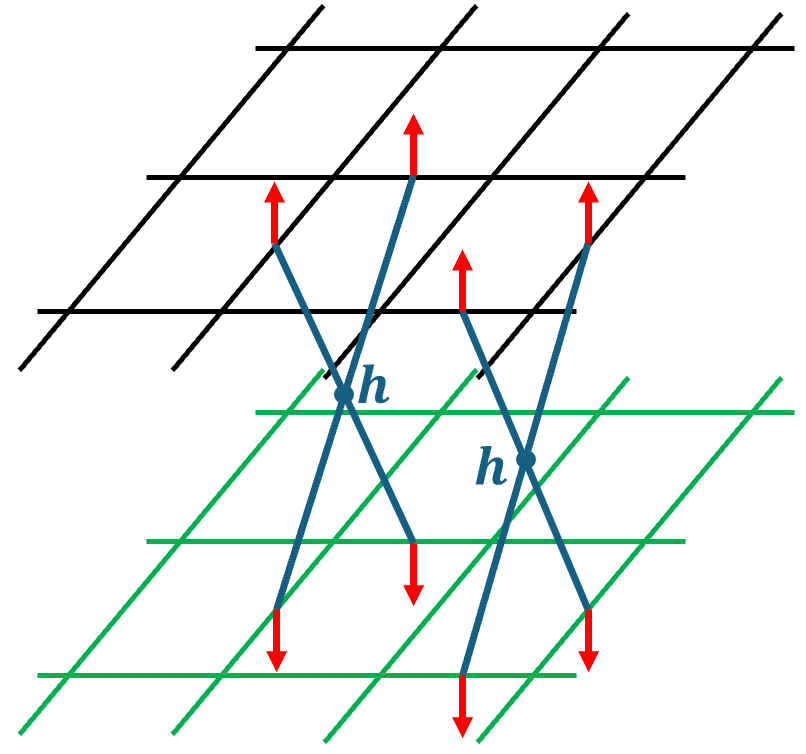}
        \caption{}
        \label{fig:DS}
    \end{subfigure}
    \captionsetup{justification=raggedright, singlelinecheck=false}
    \caption{\footnotesize{(a) The tensor $h$ is a projector that connects corresponding links of the bra and ket layer to produce a maximally mixed density matrix in the subspace where $A_v$'s are stabilizers. (b) For noise that acts on neighboring sites, $h$ is generalized to enforce the product of eigenvalues of connected links to be trivial.}}
\end{figure}

The tensor network representation of \eqref{eq:1+Av} can be constructed from two layers of pure-state tensor networks, which correspond to the bra and ket of the density matrix. In a similar spirit to \cite{guo2024locallypurifieddensityoperators}, we attach to each $g$ tensor in each layer an ancillary leg $\kappa$ that naturally extends its projective definition:
\begin{equation}
    g^j_{\alpha\beta}\rightarrow g^j_{\alpha\beta\kappa}=\delta_{\alpha,\beta,\kappa,j},
\end{equation}
where again $\delta_{\alpha,\beta,\kappa,j}=\delta_{\alpha\beta}\delta_{\beta \kappa}\delta_{\kappa j}$ is a product of Kronecker deltas.
Next, we contract the ancillary degrees of freedom with a tensor $h_{\kappa\kappa'}=\delta_{\kappa,\kappa'}$ that connects the bra and ket layers\footnote{The use of $h$ rather than direct contraction may seem redundant, but will be needed for the more general cases treated later.},  see Fig. \ref{fig:1plusBP}.

Contracting over the non-physical indices, we arrive at:
\begin{align}
\rho &=\frac{1}{2^{M/2+1}} \sum_{\{j,j'\}} \left[ \prod_v T^v(T^v)^* \prod_l g^{j_l} \left(g^{j'_{l}}\right)^* h^l \right]\notag \\
&\quad \times \;\; \ket{j_1,j_2,\ldots,j_M} \bra{j'_1,j'_2,\ldots,j'_M}.
\label{eq:flux_Z2}
\end{align} 
In this simple case, contracting $h$ results in a convex sum of closed loops formed by $j=1$ links, each of which is stabilized by all $A_v$, which shows that this representation is equivalent to (\ref{eq:1+Av}).

The RG is practically identical to the pure state case. The norm $\braket{\psi|\psi}$ is replaced with the trace of the density matrix $\text{Tr}(\rho)$, 
and the RG then proceeds as in sec. \ref{sec:rev}. 
We remark that the tensor network representation \eqref{eq:flux_Z2} of the mixed state is both an RG fixed point (invariant under renormalization) and a decoherence fixed point (maximally decohered state). For all the examples we have, these two understandings can be used interchangeably. However, to be more precise,  whenever talking about fixed points, we refer to the former.

\subsection{General protocol} \label{sec:General Protocol}

We now demonstrate a general procedure for obtaining the tensor network representation of the strongly decohered state of the $\Z_N$ topological order subject to generalized Pauli noise. Consider decoherence modeled by the following quantum channel:
\begin{equation}
    \mathcal{N}\left[\rho_0 \right]=\circ_l\;\mathcal{N}_l ,\quad \mathcal{N}_l\left[(\cdot)\right]=\sum_{a}K_{a,l}(\cdot) K^\dagger_{a,l},
\end{equation}
with the requirement that $\sum_{a} K^\dagger_{a,l}K_{a,l}=I$. Each individual $K_{a,l}$ is supported in some local region $R_l$ near $l$, and is a product of generalized Pauli operators defined in Eq. \ref{eq:general Pauli}:
\begin{equation}
    K_{a,l}=\frac{1}{n_a}\prod_{k\in R_l}\mathcal{P}_{a,k},
\end{equation}
where $n_a$ is a normalization constant. 
Since these operators commute up to a phase, and since each term in the channel consists of both Pauli strings $K$ and $K^\dagger$, different local channels $\mathcal{N}_l$ commute.

Below we will employ the Choi–Jamiołkowski isomorphism \cite{CHOI1975285,Jamiokowski1972LinearTW} to map the initial density matrix $\rho_0=\ket{\psi}\bra{\psi}$ to a state $|\rho_0\rangle\!\rangle=\ket{\psi}\otimes\ket{\psi^*}$ in the doubled Hilbert space $\mathcal{H}\otimes\mathcal{H}^*$.The tensor network representation for the post-channel Choi state will first be constructed, and from there, the tensor network representation of the original density matrix will be developed. In the Choi state, the channel maps to a summation of bra-ket paired operators acting on the doubled space:
\begin{equation}
    \circ_l\sum_{a}K_{a,l}(\cdot) K^\dagger_{a,l}\rightarrow\prod_l\sum_{a} K_{a,l}\overbar{K}_{a,l}\equiv \prod_l\sum_{a} \widetilde{K}_{a,l}. 
\label{eq:operator_doubled}
\end{equation} 
The bar indicates complex-conjugated action on $\mathcal{H}^*$. We will generally use a tilde to denote the tensor product of an operator's action on the bra and ket layers in the doubled space.

We will consider the case where the strong decoherence channel corresponds to projective measurements of anyon-creation operators with random measurement outcomes. The anyonic excitations in the Choi space have trivial self- and mutual- bosonic statistics since the corresponding $\tilde{K}$ operators commute. Specifically,  the different Kraus operators $K_{a,l}$ will be different powers of some Pauli string such that their sum $\sum_a\widetilde{K}_{a,l}=P_l(c)$ is a projector $P_l(c)$ onto an eigenvalue $c$ of this Pauli string defined in region $R_l$. This can be further decomposed:
\begin{equation} \label{eq:general projector}
    P_l(c)=\sum_{\{c_i\}}^{\prod_i c_i=c} \prod_{i\in R_l} P_{l,i} (c_i) ,
\end{equation}
where the summand is a product of projectors $P_{l,i}$, each projecting the qudit on one link into eigenvalue $c_i$. The summation is over the projectors such that $\prod_i c_i = c$.

Since the channels commute, so do the operators in \eqref{eq:operator_doubled}. Consequently, one can find a basis that simultaneously diagonalizes all components of the paired Kraus operators $\widetilde{K}$. Let $\widetilde{U}^{(x,z)}_{j,j'}$ denote the on-site basis transformation from the shared $\widetilde{Z}$ and $\widetilde{X}$ basis to the original $Z$ basis, where $x,z$ label the powers of the eigenvalues of $\widetilde{X},\widetilde{Z}$. For example, $\widetilde{X}\ket{(x,z)}=\omega_N^x\ket{(x,z)}$.

\begin{figure}
    \centering
    \includegraphics[width=0.65\columnwidth]{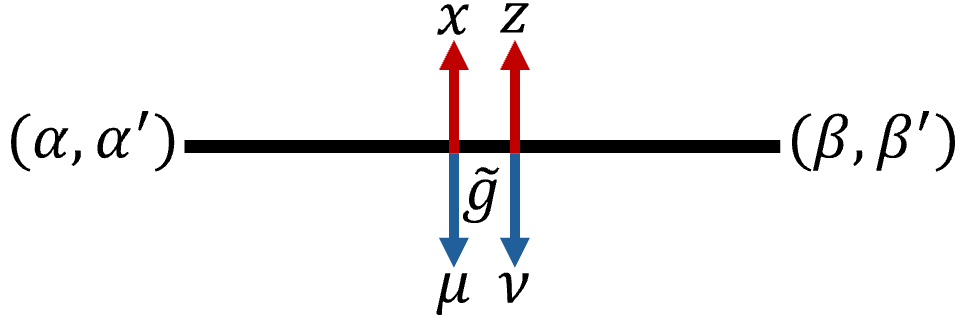}
    \captionsetup{justification=raggedright, singlelinecheck=false}
    \caption{\footnotesize{The general form of the on-site tensor $\widetilde{g}$ after diagonalizing the doubled Hilbert space Pauli strings. The physical degrees of freedom $x$ and $z$ no longer separately refer to the bra and ket layer, as $j$ and $j'$ did prior to diagonalization.}}
    \label{fig:general g}
\end{figure}

Prior to the application of the channel, the tensor network representation of the pure state in the doubled space is:
\begin{align}\label{eq:general pure double}
    &\ket{\rho_0}\!\rangle\propto\sum_{\{x,z\}} \left[ \prod_v\widetilde{T}^v \prod_l\widetilde{g}^{(x_l,z_l)} \right] \ket{\{x,z\}},
\end{align}
(step (i) in Fig. \ref{fig:flowchart}) where we have defined $\widetilde{T}^v=T^vT^{*v}$ at each vertex $v$, and:
\begin{equation} \label{eq:general g tilde}
\widetilde{g}^{(x,z)}_{\alpha\beta\alpha'\beta'}=\sum_{j,j'}\widetilde{U}^{(x,z)}_{j,j'}g^j_{\alpha\beta}\left(g^{j'}_{\alpha'\beta'}\right)^*.
\end{equation}
The tensor $\widetilde{g}$ is depicted in Fig. \ref{fig:general g}.
As before, we first extend the $\widetilde{g}$ tensor to include additional virtual indices $\mu, \nu$:
\begin{equation}
    \widetilde{g}^{(x,z)}_{\alpha\beta\alpha'\beta'} \rightarrow \widetilde{g}^{(x,z)}_{\alpha\beta\alpha'\beta',\mu\nu}=\delta_{x,\mu}\delta_{z,\nu}\widetilde{g}^{(x,z)}_{\alpha\beta\alpha'\beta'}.
\label{eq:Zn_extension}
\end{equation}
(step (ii) in Fig. \ref{fig:flowchart}). The enforcement of the product from the projector (\ref{eq:general projector}) is then set by the connecting tensor $h_l$:
\begin{equation}
    h^l_{\mu_{l_1}\nu_{l_2}...}=\begin{cases}
        1&(\mu_{l_1}+\nu_{l_2}+\cdots)\text{ mod }N=q_l,\\0&(\mu_{l_1}+\nu_{l_2}+\cdots)\text{ mod }N\neq q_l,
    \end{cases}
\label{eq:general_h}
\end{equation}
where $\omega_N^{q_l}=c_l$, which connects to $\widetilde{g}$. Including $h^l$ in the TN results in the desired maximally decohered state:
\begin{equation}
    \ket{\rho}\!\rangle=\frac{1}{C}\sum_{\{x,z\}} \left[ \prod_v\widetilde{T}^v \prod_l\widetilde{g}^{(x_l,z_l)}h^l \right] \ket{\{x,z\}},
\end{equation}
(step (iii) of Fig. \ref{fig:flowchart}) with $C=\langle\!\braket{\rho|\rho}\!\rangle$ a normalization constant. 
The tensor network of the post-channel density matrix can then be found by returning to the separated on-site basis and reversing the isomorphism:
\begin{align}
&\boxed{\rho=\frac{1}{C}\sum_{\{ j,j' \}} \left[ \prod_v\widetilde{T}^v \prod_l\widetilde{\mathfrak{g}}^{(j_l,j'_l)}h^l \right]\ket{\{j \}}\bra{\{j'\}},\notag}\\
    & \widetilde{\mathfrak{g}}^{(j,j')}_{\alpha\beta\alpha'\beta',\mu\nu}=\widetilde{g}^{(\mu,\nu)}_{\alpha\beta\alpha'\beta'}\left( \widetilde{U}^\dagger \right)^{(\mu,\nu)}_{j,j'}.
 \label{eq: physical density matrix}
\end{align}

(step (iv) in Fig. \ref{fig:flowchart}). This is one of the main results of this paper. We will use representative examples below to illustrate this formula.

\subsection{Example of decoherence in \texorpdfstring{$\Z_N$}{Z-N} topological order} \label{sec:examples}

We consider in this section two specific general forms of decoherence channels: $X^a$-type and $X^a\!Z^b$-type channels.

\subsubsection{Pure-flux decoherence}
Consider here a channel where each Kraus operator only acts on a single link, of the form:
\begin{equation} \label{eq:x channel}
    \mathcal{N}_l(\rho)=\frac{1}{N}\sum_{j=0}^{N-1}(X_l^a)^j\rho(X_l^a)^{-j}.
\end{equation}
The single-link support of the Kraus operators makes the doubled space approach unnecessary.
For $N=2$, $a=1$, this reduces to the case covered in sec. \ref{sec:1+Av}. Note that the number of decohered anyons is dependent on the greatest common divisor of $N$ and $a$, $\text{GCD}(N,a)$. For example, in $\Z_4$ topological order, if we take $a=1$ or $3$, all fluxes will be proliferated. However if we take $a=2$, the elementary flux will still survive the channel.

The channel in \eqref{eq:x channel} projects eigenvalues of $X^a$ in the bra and ket layers to be equal.  We extend the definition of $g$ as in \ref{sec:1+Av}, $g^j_{\alpha\beta}\rightarrow g^j_{\alpha\beta\kappa}$, and include the connecting tensor $h_{\kappa\kappa'}$ now defined as:
\begin{equation}
    h_{\kappa\kappa'}=\begin{cases}
        1&a(\kappa-\kappa')\text{ mod }N=0\\0&a(\kappa-\kappa')\text{ mod }N\neq0.
    \end{cases}
\end{equation}
The resulting tensor network has the same form as in Fig. \ref{fig:1plusBP}.

The RG is performed by contracting local physical indices as in $\text{Tr}(\rho)$. The density matrix may have off-diagonal elements in this basis when $\text{GCD}(N,a)\neq1$, but only the diagonal elements are relevant to the RG, such that the process is independent of $a$. The result is the same as that in sec. \ref{sec:1+Av RG} and again an RG fixed-point, with the constant of proportionality being $N$: $\widetilde{\textbf{T}}=N\widetilde{T}$.

\subsubsection{Dyonic decoherence} \label{sec:XZ-type channel}

An important channel type to consider are those of the form $X^a_lZ^b_{l-\textbf{a}}$, where $\textbf{a}$ is the constant vector defined in Fig. \ref{fig:DS stabilizers}; they represent the most general form for the decoherence of an arbitrary individual anyon $m^ae^b$ in the $\Z_N$ topological phase. The general form of the projector in \eqref{eq:general projector} that makes up the channel is:
\begin{equation} \label{eq:double projector}
\begin{aligned}
    &P_{(\omega_N^a)^x,(\omega_N^b)^z}(\widetilde{X}^a_l\widetilde{Z}^b_{l-\textbf{a}})=P_{(\omega_N^a)^x}(\widetilde{X}^a_l)P_{(\omega_N^b)^z}(\widetilde{Z}^b_{l-\textbf{a}})\\&=\frac{1}{N^2}\sum_{i,j=0}^{N-1}(\omega_N^a)^{-ix}(\omega_N^b)^{-jz} \left(\widetilde{X}^a_l\right)^i\left(\widetilde{Z}^b_{l-\textbf{a}}\right)^j.
\end{aligned}
\end{equation}
Here again $\omega_N=e^{\mathrm{i}2\pi/N}$. The number of unique eigenvalues for $\widetilde{X}^a$ will depend on $\text{GCD}(N,a)$, e.g. if $\text{GCD}(N,a)=1$, then there are $N$ unique eigenvalues (similarly for $\widetilde{Z}^b$). The channel is then expressible as the following operator acting on the Choi state:
\begin{equation} \label{eq: general choi projector}
    \mathcal{N}^\text{Choi}_l=\sum_{x,z}^{\omega_N^{x+z}=1}P_{(\omega_N^a)^x,(\omega_N^b)^z}(\widetilde{X}^a_l\widetilde{Z}^b_{l-\textbf{a}}).
\end{equation}
After extending the $\widetilde{g}$ as in \eqref{eq:Zn_extension}, we add the connecting tensor defined as:
\begin{equation} \label{eq:most general h}
        h_{\mu_l\nu_{l-\textbf{a}}}=\begin{cases}
        1 & (a\mu_l+b\nu_{l-\textbf{a}})\text{ mod }N=0,\\0&(a\mu_l+b\nu_{l-\textbf{a}})\text{ mod }N\neq0,
    \end{cases}
\end{equation}
which is a special case of \eqref{eq:general_h}. The tensor network representation of the maximally decohered density matrix is then \eqref{eq: physical density matrix} with this connecting tensor.

Next we discuss the behavior of this tensor network under renormalization. The trace of the density matrix corresponds in the doubled state to:
\begin{equation}
    \text{Tr}(\rho) =C\langle\!\braket{I|\rho}\!\rangle,
\end{equation}
where $\ket{I}\!\rangle\propto \sum_j |\{j\}\rangle |\{j^*\}\rangle$ and $C$ is chosen such that $\text{Tr}(\rho)=1$. Using the Choi state (\ref{eq:general pure double}) and including the connecting tensor (\ref{eq:most general h}) gives:
\begin{equation}
    \sum_{\{j,x,z\}}\left[ \prod_v\widetilde{T}^v \prod_l\widetilde{g}^{(x_l,z_l)}h^l\right] \prod_k\left(\widetilde{U}^{(x_k,z_k)}_{j_k,j_k}\right)^\dagger.
\end{equation}
The summation $\sum_{j_k}\left(\widetilde{U}^{(x_k,z_k)}_{j_k,j_k}\right)^\dagger$ are evaluated as follows: For any basis vector $\ket{x,z}$, its $\ket{j,j}$ components can be expanded as $\delta_{z,0}\sum_j\omega_N^{-jx}\ket{j,j}$. Each term in this sum has a $\widetilde{Z}$ eigenvalue of 1, i.e. $z=0$ (since $\widetilde{Z}=Z\otimes \overbar{Z}$ and $\overbar{Z}$ acts as the complex conjugate of $Z$ on the bra space). The sum under consideration is then just $\sum_j\omega_N^{-jx}=N\delta_{x,0}$. Then, for each $k$:
\begin{equation}
    \sum_{j_k}\left(\widetilde{U}^{(x_k,z_k)}_{j_k,j_k}\right)^\dagger=N\delta_{x_k,z_k,0}.
\end{equation}
Expanding $\widetilde{g}$ using Eq. (\ref{eq:general g tilde}), we have $\text{tr}(\rho)=\left[ \prod_v\widetilde{T} \prod_l\delta  \right],$
with $\delta$ enforcing the four virtual indices on each link to be equal (Fig. \ref{fig:general g} without the physical indices). The RG then proceeds in the same way as clarified in sec. \ref{sec:1+Av RG}. This is independent of the specific values of $a$ and $b$; the trace has allowed immediate evaluation of the connectors $h$ which contain all information about the channel.

More generally, one can show the invariance of $\text{Tr}(\rho^n)$ under the TERG procedure, which requires reformulating the TN in an error basis. details of the procedure are in Appendix \ref{sec:Tr rho n}. For the non-fixed point case, a single course-graining step in the procedure produces a rescaled connecting tensor $h$ and vertex tensor $T$ of the tensor network when Kraus operators of the channel strictly act on a single link, such as pure $e$ or $m$ decoherence. Beyond such special cases, a course-graining step 
will result in a tensor that couples the degrees of freedom of the bra and ket layers in a way that cannot be redistributed over the links again. 
Therefore, the general procedure outlined in Fig. \ref{fig:RGprocedure} will not go through directly.

\subsubsection{Fermion channel in the $\Z_2$ toric code}

As a special example of the previous subsection, here we visit the simplest case of non-bosonic decoherence: maximally decohering the fermion $f=em$ in the $\Z_2$ toric code \cite{Wang_2025,PRXQuantum.6.010313,PRXQuantum.6.010315,bao2023mixedstatetopologicalordererrorfield}. The local channel acts as:
\begin{equation}
\mathcal{N}_l \left[ (\cdot) \right]=\frac{1}{2}\left( (\cdot)+X_lZ_{l-\textbf{a}}(\cdot) X_lZ_{l-\textbf{a}} \right).
\label{eq:fermion channel}
\end{equation}
In the Choi state,  the projector form of $\mathcal{N}^\text{Choi}_l$ corresponds to the case of $N=2,\;a=b=1$ in (\ref{eq: general choi projector}) as expected:
\begin{equation}    \mathcal{N}^\text{Choi}_l=P_{-1,-1}+P_{1,1}=\frac{1}{2}\left(1+\widetilde{X}_l\widetilde{Z}_{l-\textbf{a}}\right),
\end{equation}
The basis elements are represented by $\ket{x,z}$, with $x,z\in\{0,1\}$, which are simply the four Bell states entangling a link between the bra and ket layers. The connector $h_{\mu_l\nu_{l-\textbf{a}}}$, as defined in (\ref{eq:most general h}), completes the tensor network representation of the Choi state:
\begin{equation}
    \ket{\rho}\!\rangle\propto\sum_{\{x,z\}}\left[ \prod_v\widetilde{T}^v\prod_l \widetilde{g}^{(x_l,z_l)}h^l \right]\ket{\{x,z\}}.
\end{equation}
The tensor network representation of the density matrix is then \eqref{eq: physical density matrix}, where the tensors are summarized as follows\footnote{In the matrix representation of $\widetilde{U}$, the different rows (columns) correspond to different values of $\{j,j'\}$ $(\{x,z\})$, with the following ordering from top (left) to bottom (right): $(\{0,0\},\{1,0\},\{0,1\},\{1,1\})$.}:
\begin{gather}
    \widetilde{U}=\frac{1}{\sqrt{2}}\begin{pmatrix}
        1&1&0&0\\0&0&1&-1\\0&0&1&1\\1&-1&0&0
    \end{pmatrix},\\
    \widetilde{g}^{(x,z)}_{\alpha\beta\alpha'\beta',\mu\nu}=\delta_{x,\mu}\delta_{z,\nu}\delta_{\alpha,\beta}\delta_{\alpha',\beta'}\widetilde{U}^{(x,z)}_{\alpha,\alpha'},\\ \label{eq:h_fermion}
    h_{\mu_l\nu_{l-\textbf{a}}}=\begin{cases}
        1 & (\mu_l+\nu_{l-\textbf{a}})\text{ mod }2=0,\\0&(\mu_l+\nu_{l-\textbf{a}})\text{ mod }2\neq0.
    \end{cases}
\end{gather}

There is also an alternative representation of this state using a fermionic tensor network description, see Appendix \ref{sec:bosonization}.

\subsubsection{Decohering \texorpdfstring{$e^2m^2$}{e2m2} from \texorpdfstring{$\Z_4$}{Z-4} toric code}\label{sec:DS mixed}

Consider the $\Z_4$ quantum double with maximal decoherence of the dyon $e^2m^2$:  
\begin{equation} \label{eq:DS channel}
    \mathcal{N}_l \left[ (\cdot) \right]=\frac{1}{2}\left( (\cdot)+X^2_lZ^2_{l-\textbf{a}}(\cdot) X^2_lZ^2_{l-\textbf{a}} \right).
\end{equation}
Since $e^2m^2$ is a self-boson, the bases of $\mathcal{H}_l$ and $\mathcal{H}^*_l$ can separately be taken to diagonalize $X^2_l$ and $Z^2_l$, which allows us to skip the doubled representation altogether. 
The projector form of the channel (\ref{eq:DS channel}) is:
\begin{equation}
    \mathcal{N}_l\left[(\cdot)\right]=\sum_{x,z,x'z'}\!\!\!\!\!^{'}P_{(-1)^x,(-1)^z}(\cdot)P^\dagger_{(-1)^{x'},(-1)^{z'}},
\end{equation}
with the sum restriction $\sum'$ indicating $(-1)^{-x-z+x'+z'}=1$. Here,$\:x,z\in\{0,1\}$ are powers of eigenvalues for $X^2$ and $Z^2$ as in sec. \ref{sec:DS pure}: $X^2\ket{x,z}=(-1)^x\ket{x,z},\quad Z^2\ket{x,z}=(-1)^z\ket{x,z}$.

The basis change can be done separately in each layer, leading to:
\begin{equation}
    g^{(x,z)}_{\alpha\beta}=\sum_j U^{(x,z)}_jg^j_{\alpha\beta}.
\end{equation}
After extending $g$ with ancillae $g^{(x,z)}_{\alpha\beta}\rightarrow g^{(x,z)}_{\alpha\beta,\mu\nu}$ as in equation (\ref{eq:g extend pure}), the connector $h$ projects onto states where the product of eigenvalues of $X^2_l$, $Z^2_{l-\textbf{a}}$, $\overbar{X}^2_l$, and $\overbar{Z}^2_{l-\textbf{a}}$ is 1:
\begin{equation}
    h_{\mu_l\nu_{l-\textbf{a}}\mu'_l\nu'_{l-\textbf{a}}}=\begin{cases}
        1&(\mu_l+\nu_{l-\textbf{a}}-\mu'_l-\nu'_{l-\textbf{a}})\text{ mod }2=0,\\0&(\mu_l+\nu_{l-\textbf{a}}-\mu'_l-\nu'_{l-\textbf{a}})\text{ mod }2\neq0.
    \end{cases}
\label{eq:h_DS}
\end{equation}
The resulting state is:
\begin{align}\label{eq:DS mixed}
    \rho=&\frac{1}{4^{M/2+1}}\sum\left[ \prod_v\widetilde{T}^v\prod_l g^{(x_l,z_l)} g^{(x'_l,z'_l)}h^l \right] \notag\\&\times\ket{\{(x,z\}}\bra{\{(x',z')\}}
\end{align}
The TN is shown in Fig. \ref{fig:DS}. This state is stabilized by $A_v^\text{DS}$ and $B_p^\text{DS}$ (Fig. \ref{fig:DS stabilizers}) and is maximally mixed in the resulting subspace:
\begin{equation} \label{eq:rho DS}
    \rho=\frac{1}{2^{M/2+3}}\prod_v\frac{1+A_v^\text{DS}+(A_v^\text{DS})^2+(A_v^\text{DS})^3}{4}\prod_p\frac{1+B_p^{\text{DS}}}{2}.
\end{equation}
This maximal mixture is among the pure states which (1) are all stabilized by the $A_v^{\text{DS}}$ and the $B_p^2$ operators, and (2) are stabilized by $X_l^2 Z_{l-\bm{a}}^2$ for a set of links $l\in S$ on the square lattice, and stabilized by $-X_{k}^2 Z_{k-\bm{a}}^2$ on the remaining links $k\notin S$. This set $S$ must satisfy the constraint that around each vertex, there are an even number of links $l$ that belong to $S$. This constraint comes from the fact that $\prod_{\partial l\ni v} X_{l}^2 Z_{l-\bm{a}}^2=(A_v^{\text{DS}})^2.$
It is easy to check that these pure states make up the basis in the $A_v^\text{DS}$,$B_p^\text{DS}$-stabilized subspace, and that $\rho$ in (\ref{eq:rho DS}) is the equal weight convex sum of said states, matching the result given in \cite{PRXQuantum.6.010315}. At each $l\in S$, one can also easily write down the pure-state tensor network by replacing $h$ as:
\begin{equation}
    h^\text{defect}_{\mu_l\nu_{l-\textbf{a}}}=\begin{cases}
        1 & (\mu_l+\nu_{l-\textbf{a}})\text{ mod }2\neq0,\\0&(\mu_l+\nu_{l-\textbf{a}})\text{ mod }2=0.
    \end{cases}
\end{equation}
Finally we comment on the RG, which proceeds as follows: We start by expressing the TN in a form closer to that of the general procedure, which is more suitable for the RG. 
Decompose $h_{\mu_l\nu_{l-\textbf{a}}\mu'_l\nu'_{l-\textbf{a}}}$ in equation \eqref{eq:h_DS} into a product of two tensors:
\begin{equation} \label{eq:h DS split}
\begin{aligned}
    h_{\mu_l\nu_{l-\textbf{a}}\mu'_l\nu'_{l-\textbf{a}}}&=r_{\mu_l\mu'_l\kappa}r_{\nu_{l-\textbf{a}}\nu'_{l-\textbf{a}}\kappa},\\r_{\mu_l\mu'_l\kappa}&=\delta_{\mu_l,\mu'_l,\kappa}.
\end{aligned}
\end{equation}

\tikzmath{\shiftx=0.1;\shifty=0.4;\shiftxx=0;\shiftyy=7;}
\begin{figure}
\centering
\begin{subfigure}[b]{0.45\columnwidth}
\scalebox{0.35}{
\begin{tikzpicture}
  \foreach \x in {-4,-2,0,2,4} \draw[Green,ultra thick,shift={(\shiftxx,-\shiftyy)}] (-4+0.4*\x,\x)--(4+0.4*\x,\x);
  \foreach \y in {-4,-2,0,2,4} \draw[Green,ultra thick,shift={(\shiftxx,-\shiftyy)}] (\y-0.4*4,-4)--(\y+0.4*4,4);
   \draw[NavyBlue,ultra thick] (0+0.4*-2+1,-2)--(0+2+0.4*-2+0.4+\shiftxx,-2+1-\shiftyy);
   \draw[NavyBlue,ultra thick] (0+0.4*-2+1+\shiftxx,-2-\shiftyy)--(0+2+0.4*-2+0.4,-2+1);
   \draw[NavyBlue,ultra thick] (-4+0.4*0+1,0)--(-4+2+0.4*0+0.4+\shiftxx,0+1-\shiftyy);
   \draw[NavyBlue,ultra thick] (-4+0.4*0+1+\shiftxx,0-\shiftyy)--(-2+0.4*0+0.4,0+1);
   \filldraw[NavyBlue] ({0.5*(0+0.4*-2+1+0+2+0.4*-2+0.4+\shiftxx)},{0.5*(-2-2+1-\shiftyy)}) circle [radius=4pt] node[below right=2pt] {\LARGE $\textit{\textbf{h}}$};
   \filldraw[NavyBlue] ({0.5*(-4+0.4*0+1-4+2+0.4*0+0.4+\shiftxx)},{0.5*(0+1-\shiftyy)}) circle [radius=4pt] node[above right=2pt] {\LARGE $\textit{\textbf{h}}$};
  \foreach \x in {-4,-2,0,2,4} \draw[black,ultra thick] (-4+0.4*\x,\x)--(4+0.4*\x,\x);
  \foreach \y in {-4,-2,0,2,4} \draw[black,ultra thick] (\y-0.4*4,-4)--(\y+0.4*4,4);
  \foreach \x/\y in {-4/0,0/-2} {
  \draw[WildStrawberry,line width = 2pt,->,>=Stealth] (\x+0.4*\y+1,\y)--(\x+0.4*\y+1,\y+0.8);
  \draw[WildStrawberry,line width = 2pt,->,>=Stealth] (\x+0.4*\y+0.4+2,\y+1)--(\x+0.4*\y+0.4+2,\y+1+.8);
  \draw[WildStrawberry,line width = 2pt,->,>=Stealth] (\x+0.4*\y+1+\shiftxx,\y-\shiftyy)--(\x+0.4*\y+1+\shiftxx,\y-.8-\shiftyy);
  \draw[WildStrawberry,line width = 2pt,->,>=Stealth] (\x+0.4*\y+0.4+2+\shiftxx,\y+1-\shiftyy)--(\x+0.4*\y+0.4+2+\shiftxx,\y+1-.8-\shiftyy);
  }
\end{tikzpicture}
}
\caption{}
\end{subfigure}
\begin{subfigure}[b]{0.45\columnwidth}
\scalebox{0.35}{
\begin{tikzpicture}
  \coordinate (h1) at ({0.5*(-4+0.4*0+1-4+2+0.4*0+0.4+\shiftxx)},{0.5*(0+1-\shiftyy)});
  \coordinate (h2) at ({0.5*(0+0.4*-2+1+0+2+0.4*-2+0.4+\shiftxx)},{0.5*(-2-2+1-\shiftyy)});
  \coordinate (splitshift) at (0.25,0.15);
  \coordinate (r1h1) at ($(h1)-(splitshift)$);
  \coordinate (r2h1) at ($(h1)+(splitshift)$);
  \coordinate (r1h2) at ($(h2)-(splitshift)$);
  \coordinate (r2h2) at ($(h2)+(splitshift)$);
  \foreach \x in {-4,-2,0,2,4} \draw[Green,ultra thick,shift={(\shiftxx,-\shiftyy)}] (-4+0.4*\x,\x)--(4+0.4*\x,\x);
  \foreach \y in {-4,-2,0,2,4} \draw[Green,ultra thick,shift={(\shiftxx,-\shiftyy)}] (\y-0.4*4,-4)--(\y+0.4*4,4);
   \draw[NavyBlue,ultra thick] (0+0.4*-2+1,-2)--(r1h2);
   \draw[NavyBlue,ultra thick] (r2h2)--(0+2+0.4*-2+0.4+\shiftxx,-2+1-\shiftyy);
   \draw[NavyBlue,ultra thick] (0+0.4*-2+1+\shiftxx,-2-\shiftyy)--(r1h2);
   \draw[NavyBlue,ultra thick] (r2h2)--(0+2+0.4*-2+0.4,-2+1);
   \draw[NavyBlue,ultra thick] (-4+0.4*0+1,0)--(r1h1);
   \draw[NavyBlue,ultra thick] (r2h1)--(-4+2+0.4*0+0.4+\shiftxx,0+1-\shiftyy);
   \draw[NavyBlue,ultra thick] (-4+0.4*0+1+\shiftxx,0-\shiftyy)--(r1h1);
   \draw[NavyBlue,ultra thick] (r2h1)--(-2+0.4*0+0.4,0+1);
   \filldraw[RoyalPurple] (r1h2) circle [radius=3pt] node[left=2pt] {\LARGE $\textit{\textbf{r}}$};
   \filldraw[RoyalPurple] (r2h2) circle [radius=3pt] node[right=2pt] {\LARGE $\textit{\textbf{r}}$};
   \filldraw[RoyalPurple] (r1h1) circle [radius=3pt] node[left=2pt] {\LARGE $\textit{\textbf{r}}$};
   \filldraw[RoyalPurple] (r2h1) circle [radius=3pt] node[right=2pt] {\LARGE $\textit{\textbf{r}}$};
   \draw[RoyalPurple,ultra thick] (r1h1)--(r2h1);
   \draw[RoyalPurple,ultra thick] (r1h2)--(r2h2);
  \foreach \x in {-4,-2,0,2,4} \draw[black,ultra thick] (-4+0.4*\x,\x)--(4+0.4*\x,\x);
  \foreach \y in {-4,-2,0,2,4} \draw[black,ultra thick] (\y-0.4*4,-4)--(\y+0.4*4,4);
  \foreach \x/\y in {-4/0,0/-2} {
  \draw[WildStrawberry,line width = 2pt,->,>=Stealth] (\x+0.4*\y+1,\y)--(\x+0.4*\y+1,\y+0.8);
  \draw[WildStrawberry,line width = 2pt,->,>=Stealth] (\x+0.4*\y+0.4+2,\y+1)--(\x+0.4*\y+0.4+2,\y+1+.8);
  \draw[WildStrawberry,line width = 2pt,->,>=Stealth] (\x+0.4*\y+1+\shiftxx,\y-\shiftyy)--(\x+0.4*\y+1+\shiftxx,\y-.8-\shiftyy);
  \draw[WildStrawberry,line width = 2pt,->,>=Stealth] (\x+0.4*\y+0.4+2+\shiftxx,\y+1-\shiftyy)--(\x+0.4*\y+0.4+2+\shiftxx,\y+1-.8-\shiftyy);
  }
\end{tikzpicture}
}
\caption{}
\end{subfigure}
\begin{subfigure}[b]{\columnwidth}
\scalebox{0.35}{
\begin{tikzpicture}
  \foreach \x in {-4,-2,0,2} \foreach \y in {-4,-2,0,2,4} {
  \draw[black,ultra thick] (\x+0.4*\y,\y)--(\x+0.4*\y+1+0.5*\shiftx,\y-0.5*\shifty);
  \draw[black,ultra thick] (\x+0.4*\y+1+0.5*\shiftx,\y-0.5*\shifty)--(\x+2+0.4*\y,\y);
  \draw[Green,ultra thick] (\x+0.4*\y+\shiftx,\y-\shifty)--(\x+0.4*\y+1+0.5*\shiftx,\y-0.5*\shifty);
  \draw[Green,ultra thick] (\x+0.4*\y+1+0.5*\shiftx,\y-0.5*\shifty)--(\x+2+0.4*\y+\shiftx,\y-\shifty);
  \draw[black,ultra thick] (\y+0.4*\x,\x)--(0.5*\y+0.5*0.4*\x+0.5*\y+0.5*0.4*\x+0.5*0.8+0.5*\shiftx,0.5*\x+0.5*\x+0.5*2-0.5*\shifty);
  \draw[black,ultra thick] (0.5*\y+0.5*0.4*\x+0.5*\y+0.5*0.4*\x+0.5*0.8+0.5*\shiftx,0.5*\x+0.5*\x+0.5*2-0.5*\shifty)--(\y+0.4*\x+0.8,\x+2);
  \draw[Green,ultra thick] (\y+0.4*\x+\shiftx,\x-\shifty)--(0.5*\y+0.5*0.4*\x+0.5*\y+0.5*0.4*\x+0.5*0.8+0.5*\shiftx,0.5*\x+0.5*\x+0.5*2-0.5*\shifty);
  \draw[Green,ultra thick] (0.5*\y+0.5*0.4*\x+0.5*\y+0.5*0.4*\x+0.5*0.8+0.5*\shiftx,0.5*\x+0.5*\x+0.5*2-0.5*\shifty)--(\y+0.4*\x+0.8+\shiftx,\x+2-\shifty);
  \filldraw[RoyalPurple] (\x+0.4*\y+1+0.5*\shiftx,\y-0.5*\shifty) circle [radius=3pt];
  \filldraw[RoyalPurple] (0.5*\y+0.5*0.4*\x+0.5*\y+0.5*0.4*\x+0.5*0.8+0.5*\shiftx,0.5*\x+0.5*\x+0.5*2-0.5*\shifty) circle [radius=3pt];
  }
  \foreach \x/\y in {-4/0,0/-2} {
   \draw[RoyalPurple,ultra thick](\x+0.4*\y+1+0.5*\shiftx,\y-0.5*\shifty)--(0.5*\x+0.5*0.4*\y+0.5*\x+0.5*0.4*\y+0.5*0.8+0.5*\shiftx+2,0.5*\y+0.5*\y+0.5*2-0.5*\shifty);
   \draw[WildStrawberry,line width = 2pt,->,>=Stealth] (\x+0.4*\y+1+0.5*\shiftx,\y-0.5*\shifty) -- (\x+0.4*\y+1+0.5*\shiftx,\y-0.5*\shifty+0.8);
   \draw[WildStrawberry,line width = 2pt,->,>=Stealth] (\x+0.4*\y+1+0.5*\shiftx,\y-0.5*\shifty) -- (\x+0.4*\y+1+0.5*\shiftx,\y-0.5*\shifty-0.8);
   \draw[WildStrawberry,line width = 2pt,->,>=Stealth] (0.5*\x+0.5*0.4*\y+0.5*\x+0.5*0.4*\y+0.5*0.8+0.5*\shiftx+2,0.5*\y+0.5*\y+0.5*2-0.5*\shifty) -- (0.5*\x+0.5*0.4*\y+0.5*\x+0.5*0.4*\y+0.5*0.8+0.5*\shiftx+2,0.5*\y+0.5*\y+0.5*2-0.5*\shifty+0.8);
   \draw[WildStrawberry,line width = 2pt,->,>=Stealth] (0.5*\x+0.5*0.4*\y+0.5*\x+0.5*0.4*\y+0.5*0.8+0.5*\shiftx+2,0.5*\y+0.5*\y+0.5*2-0.5*\shifty) -- (0.5*\x+0.5*0.4*\y+0.5*\x+0.5*0.4*\y+0.5*0.8+0.5*\shiftx+2,0.5*\y+0.5*\y+0.5*2-0.5*\shifty-0.8);
   }
\end{tikzpicture}
}
\caption{}
\end{subfigure}
\captionsetup{justification=raggedright, singlelinecheck=false}
\caption{\footnotesize{The RG procedure for the decoherence in sec. \ref{sec:DS mixed}. Only two connectors $h$ are shown for visual clarity. The $h$ are first split as into products of $r$ as in (\ref{eq:h DS split}) (b), followed by a contraction of the blue indices that leads to (c).}}
\label{fig:DS mixed RG}
\end{figure}
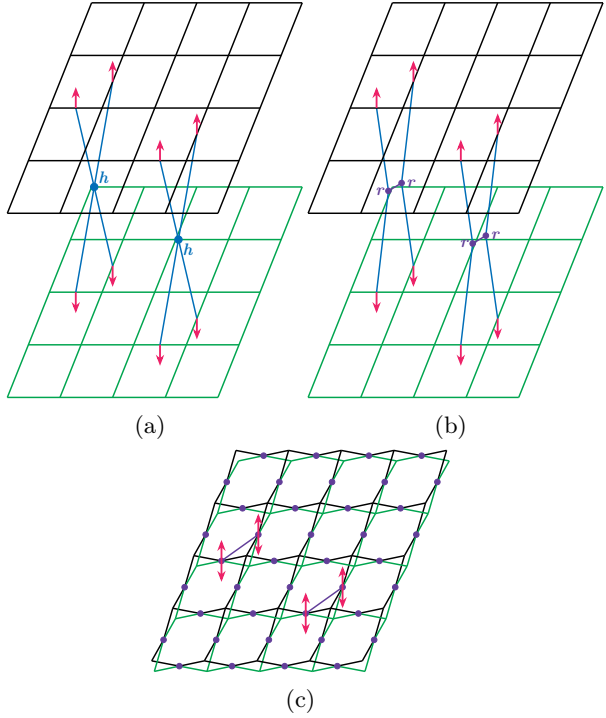
The $\mu$ and $\nu$ indices can now be contracted for each individual link, and the resulting TN appears as in Fig. \ref{fig:DS mixed RG}. From here, taking the trace will simply give $\delta_{\kappa,\kappa',0}\delta_{\alpha,\beta,\alpha'\beta'}$, i.e. the diagonal contracting tensors $r$ completely detach from the rest of the TN, and the latter reduces to the pure-state RG, as expected from sec.\ \ref{sec:XZ-type channel}. Therefore, after first contracting out $r$, the RG proceeds as in the pure case of sec. \ref{sec:1+Av RG}, as expected from the result in sec. \ref{sec:XZ-type channel}.

\section{Extensions}
\label{sec:extensions}

In this section, we go beyond the decohered two-dimensional $\Z_N$ topological order and list a few more interesting examples, including fixed-point TN representation of decohered non-Abelian $D(S_3)$ topological order (\ref{sec:S3}), pure-flux decoherence of arbitrary CSS codes (\ref{sec:CSS}) with three-dimensional toric code (\ref{sec:3D}) being a representative, and chiral topological phases (\ref{sec:chiral}).

\subsection{Decoherence in the \texorpdfstring{$D(S_3)$}{S3} topological phase} 
\label{sec:S3}

This section concerns the decoherence of a non-Abelian $D(S_3)$ topological order. The pure-state tensor network representation of quantum double with finite group  $G$ \cite{Kitaev_2003} was worked out in \cite{JMathPhys.54.012201}, which we briefly review here.

For input group $G$, the Hilbert space on each link is spanned by $\{ \ket{g},g\in G\}$. For convenience we introduce an orientation on each link as usual, and reversing the orientation corresponds to the map $\ket{g}\rightarrow\ket{-g}$. The Kitaev quantum double Hamiltonian is written with the following operations:
\begin{equation}
\begin{aligned}
    &L_+(g)\ket{g'}=\ket{gg'},& &T_+(g)\ket{g'}=\delta_{g,g'}\ket{g'},\\&L_-(g)\ket{g'}=\ket{g'g^{-1}},& &T_-(g)\ket{g'}=\delta_{g^{-1},g'}\ket{g'},
\end{aligned}
\end{equation}
i.e., $L_+$ is simply the left regular unitary representation of $G$. 
The vertex projection operators are constructed as:
\begin{equation}
    A_v=\frac{1}{|G|}\sum_{g\in G}L_+^{g_\downarrow}(g) L_+^{g_\leftarrow}(g) L_-^{g_\uparrow}(g) L_-^{g_\rightarrow}(g),
\end{equation}
where the arrows in the superscripts indicate the positions of the link with respect to the vertex $v$, i.e.:
\tikzmath{\vscale=1.25;}
\begin{equation}
    A_v \;\;\tikz[baseline=-.6ex]{\draw[black,thin] (-\vscale,0)--(\vscale,0);
    \draw[black,thin] (0,-\vscale)--(0,\vscale);
    \draw[black,very thin] (-\vscale-0.1,-\vscale-0.3)--(-\vscale-0.1,\vscale+0.3);
    \draw[black,very thin] (\vscale-0.35,-\vscale-0.3)--(\vscale+0.1,0)--(\vscale-0.35,\vscale+0.3);
    \node[isosceles triangle,draw,fill=black, minimum size=0.1cm,inner sep=0pt,rotate=0] at (-\vscale/2,0){};
    \node[isosceles triangle,draw,fill=black, minimum size=0.1cm,inner sep=0pt,rotate=0] at (\vscale/2,0){};
    \node[isosceles triangle,draw,fill=black, minimum size=0.1cm,inner sep=0pt,rotate=90] at (0,-\vscale/2){};
    \node[isosceles triangle,draw,fill=black, minimum size=0.1cm,inner sep=0pt,rotate=90] at (0,\vscale/2){};
    \node[below] at (-\vscale/2,0){$g_\leftarrow$};
    \node[above] at (\vscale/2,0){$g_\rightarrow$};
    \node[left] at (0,\vscale/2){$g_\uparrow$};
    \node[right] at (0,-\vscale/2){$g_\downarrow$};
    }
    =\sum_{g\in G} \tikz[baseline=-.6ex]{\draw[black,thin] (-\vscale,0)--(\vscale,0);
    \draw[black,thin] (0,-\vscale)--(0,\vscale);
    \draw[black,very thin] (-\vscale-0.1,-\vscale-0.3)--(-\vscale-0.1,\vscale+0.3);
    \draw[black,very thin] (\vscale-0.35,-\vscale-0.3)--(\vscale+0.1,0)--(\vscale-0.35,\vscale+0.3);
    \node[isosceles triangle,draw,fill=black, minimum size=0.1cm,inner sep=0pt,rotate=0] at (-\vscale/2,0){};
    \node[isosceles triangle,draw,fill=black, minimum size=0.1cm,inner sep=0pt,rotate=0] at (\vscale/2,0){};
    \node[isosceles triangle,draw,fill=black, minimum size=0.1cm,inner sep=0pt,rotate=90] at (0,-\vscale/2){};
    \node[isosceles triangle,draw,fill=black, minimum size=0.1cm,inner sep=0pt,rotate=90] at (0,\vscale/2){};
    \node[below] at (-\vscale/2,0){$gg_\leftarrow$};
    \node[above] at (\vscale/2,0){$g_\rightarrow g^{-1}$};
    \node[left] at (0,\vscale/2){$g_\uparrow g^{-1}$};
    \node[right] at (0,-\vscale/2){$gg_\downarrow$};
    }.
\end{equation}
The positive and negative subscripts correspond to the convention chosen for the orientation of the links in the TN; the top and right legs point out of the vertex, so their contribution to the Gauss law enforced by $A_v$ is opposite to the left and bottom legs. 

Plaquette operators $B_p$ act by projecting the product of group elements around the plaquette onto the identity:
\begin{equation}
    B_p\: \tikz[baseline=-.6ex]{\draw[black,thin] (-0.7,-0.7)--(0.7,-0.7)--(0.7,0.7)--(-0.7,0.7)--(-0.7,-0.7);
    \node at (0,-0.50) {$g_1$};
    \node at (-0.45,0) {$g_4$};
    \node at (0,0.50) {$g_3$};
    \node at (0.45,0) {$g_2$};
    \node[isosceles triangle,draw,fill=black, minimum size=0.1cm,inner sep=0pt,rotate=0] at (0,-0.7){};
    \node[isosceles triangle,draw,fill=black, minimum size=0.1cm,inner sep=0pt,rotate=0] at (0,0.7){};
    \node[isosceles triangle,draw,fill=black, minimum size=0.1cm,inner sep=0pt,rotate=90] at (0.7,0){};
    \node[isosceles triangle,draw,fill=black, minimum size=0.1cm,inner sep=0pt,rotate=90] at (-0.7,0){};}
    = \delta_{g^{-1}_4g^{-1}_3g_2g_1,e}\: \tikz[baseline=-.6ex]{\draw[black,thin] (-0.7,-0.7)--(0.7,-0.7)--(0.7,0.7)--(-0.7,0.7)--(-0.7,-0.7);
    \node at (0,-0.50) {$g_1$};
    \node at (-0.45,0) {$g_4$};
    \node at (0,0.50) {$g_3$};
    \node at (0.45,0) {$g_2$};
    \node[isosceles triangle,draw,fill=black, minimum size=0.1cm,inner sep=0pt,rotate=0] at (0,-0.7){};
    \node[isosceles triangle,draw,fill=black, minimum size=0.1cm,inner sep=0pt,rotate=0] at (0,0.7){};
    \node[isosceles triangle,draw,fill=black, minimum size=0.1cm,inner sep=0pt,rotate=90] at (0.7,0){};
    \node[isosceles triangle,draw,fill=black, minimum size=0.1cm,inner sep=0pt,rotate=90] at (-0.7,0){};}
\end{equation}
These operators all mutually commute, and the Hamiltonian is $H=-\sum_vA_v-\sum_pB_p$; then the ground state subspace is the mutual $+1$ eigenspace of all $A_v$,$B_p$'s. 

To construct the tensor network representation of the ground state, assign to each link the following tensor:
\begin{gather}
   \tikz[baseline=-.6ex]{\draw[Cerulean, thin] (-1,0)--(1,0);
   \draw[black] (0.73,0.46) arc (-25:-155:0.8);
   \draw[black] (0.73,-0.46) arc (25:155:0.8);
   \filldraw[red] (0,0) circle (1.5pt);
   \draw[red,ultra thick,-latex] (0,0)--(0.75,0) node[anchor=west]{$g$};
   \node at (0.8,0.65) {$\alpha$};
   \node at (-0.8,0.65) {$\beta$};
   \node at (0.8,-0.65) {$\delta$};
   \node at (-0.8,-0.65) {$\gamma$};
   \node[isosceles triangle,draw,fill=black, minimum size=0.1cm,inner sep=0pt,rotate=-135] at (0.58,0.25){};
   \node[isosceles triangle,draw,fill=black, minimum size=0.1cm,inner sep=0pt,rotate=131] at (-0.58,0.25){};
   \node[isosceles triangle,draw,fill=black, minimum size=0.1cm,inner sep=0pt,rotate=-45] at (0.58,-0.25){};
   \node[isosceles triangle,draw,fill=black, minimum size=0.1cm,inner sep=0pt,rotate=45] at (-0.58,-0.25){};
   }
   =
   \tikz[baseline=-.6ex]{\draw[Cerulean, thin] (0,-1)--(0,1);
   \draw[black] (0.46,-0.73) arc (-115:-250:0.8);
   \draw[black] (-0.46,0.73) arc (65.0:-65:0.8);
   \filldraw[red] (0,0) circle (1.5pt);
   \draw[red,ultra thick,-latex] (0,0)--(0,0.75) node[anchor=south]{$g^{-1}$};
   \node at (0.67,-0.77) {$\alpha$};
   \node at (0.67,0.77) {$\beta$};
   \node at (-0.67,-0.77) {$\delta$};
   \node at (-0.67,0.77) {$\gamma$};
   \node[isosceles triangle,draw,fill=black, minimum size=0.1cm,inner sep=0pt,rotate=45] at (0.28,0.6){};
   \node[isosceles triangle,draw,fill=black, minimum size=0.1cm,inner sep=0pt,rotate=134] at (0.276,-0.6){};
   \node[isosceles triangle,draw,fill=black, minimum size=0.1cm,inner sep=0pt,rotate=-45] at (-0.276,0.6){};
   \node[isosceles triangle,draw,fill=black, minimum size=0.1cm,inner sep=0pt,rotate=-135] at (-0.276,-0.6){};
   }=O^g_{\alpha\beta\gamma\delta},\notag\\O^g_{\alpha\beta\gamma\delta}=L_{+,\alpha\beta}(g)L_{+,\gamma\delta}(g^{-1}).
\end{gather}
The blue line depicts the link, either horizontal or vertical. Incoming and outgoing arcs represent row and column indices of the tensor respectively. Contraction of all the circles (one per plaquette) while summing over all physical configurations $\{g\}$ yields the ground state:
\begin{equation}
    \ket{\psi_0}=\sum_{g\in G} \left[ \prod_lO^{g_l} \right] \ket{g_1,g_2,...}.
\end{equation}
Excitations above this ground state can be expressed using operators that create anyon pairs. 
The operators that create an anyon and its conjugate are the ribbon operators $F^{(\textbf{R},C)}$, where $C$ is a choice of conjugacy class for the input group, and $\textbf{R}$ is an irreducible representation (irrep) of the centralizer of a representative element of $C$ (different choices of representatives lead to isomorphic centralizer groups). The smallest ribbon operators are \cite{chen2025universalcircuitsetusing}: 
\begin{align}
\left(F^{(\textbf{R},C)}_{l,h}\right)^{u}_{w}=\gamma\!\!\!\! \sum_{n\in Z(g_c)}\!\!\Gamma^\textbf{R}_{jj'}(n)(T_+\left(\tau_cn\tau_{c'}^{-1})\right)_l\left(L_-(c')\right)_{l-\textbf{b}},\label{eq:ribbon-h}\\
\left(F^{(\textbf{R},C)}_{l,v}\right)^{u}_{w}=\gamma\!\!\!\! \sum_{n\in Z(g_c)}\!\!\Gamma^\textbf{R}_{jj'}(n)\left(L_+(c')\right)_l\left(T_+(\tau_cn\tau_{c'}^{-1})\right)_{l+\textbf{b}}\label{eq:ribbon-v}.
\end{align}
In the equations above, $\gamma=\frac{|\textbf{R}|}{|Z(g_c)|}$, $u=(c,j)$ and $w=(c',j')$, $c,c'\in C$ label the degrees of freedom of the anyons at the two ends of the link $l$. $g_c\in C$ is a representative element of $C$, $\Gamma^\textbf{R}(n)$ is the matrix form of group element $n$ in irrep $\textbf{R}$, $h$ and $v$ correspond to horizontal and vertical ribbon paths, and $\tau_c \in G$ is defined to satisfy $\tau_cg_c\tau^{-1}_c=c$. The link $l-\textbf{b}$ is to the bottom-right of $l$. Graphically:
\begin{equation}
\adjustbox{raise=0.9cm}{$h:$ }
\begin{tikzpicture}
    \draw[Blue, ultra thick,-latex] (0.5,0)--(0,0.5);
    \node[above right] at (0.25,0.25) {\small{\textbf{\textcolor{Blue}{b}}}};
    \foreach \x in {-1,0,1} {\draw[Black, ultra thick] (\x,-1)--(\x,1); \draw[black, ultra thick] (-1,\x)--(1,\x);}
    \foreach \x in {-0.5,0.5} \foreach \y in {-1,0,1} {
    \node[isosceles triangle,draw,fill=black, minimum size=0.1cm,inner sep=0pt,rotate=0] at (\x,\y){};};
    \foreach \y in {-0.5,0.5} \foreach \x in {-1,0,1} {
    \node[isosceles triangle,draw,fill=black, minimum size=0.1cm,inner sep=0pt,rotate=90] at (\x,\y){};};
    \draw[Magenta, thick] (-1,0)--(0,0)--(-0.5,-0.5)--cycle;
    \draw[Magenta, thick] (0,0)--(-0.5,-0.5)--(0.5,-0.5)--cycle;
    \node[above] at (-0.5,0) {\small{$l$}};
    \node[below right] at (0,-0.5) {\small{$l_2$}};
\end{tikzpicture}
\adjustbox{raise=0.9cm}{ ,\quad $v:$ }
\begin{tikzpicture}
    \draw[Blue, ultra thick,-latex] (0.5,0)--(0,0.5);
    \node[above right] at (0.25,0.25) {\small{\textbf{\textcolor{Blue}{b}}}};
    \foreach \x in {-1,0,1} {\draw[Black, ultra thick] (\x,-1)--(\x,1); \draw[black, ultra thick] (-1,\x)--(1,\x);}
    \foreach \x in {-0.5,0.5} \foreach \y in {-1,0,1} {
    \node[isosceles triangle,draw,fill=black, minimum size=0.1cm,inner sep=0pt,rotate=0] at (\x,\y){};};
    \foreach \y in {-0.5,0.5} \foreach \x in {-1,0,1} {
    \node[isosceles triangle,draw,fill=black, minimum size=0.1cm,inner sep=0pt,rotate=90] at (\x,\y){};};
    \draw[Magenta, thick] (-1,0)--(-0.5,-0.5)--(-0.5,0.5)--cycle;
    \draw[Magenta, thick] (-1,0)--(-0.5,0.5)--(-1,1)--cycle;
    \node[above right] at (-0.5,0) {\small{$l$}};
    \node[right] at (-1,0.5) {\small{$l_3$}};
\end{tikzpicture}
\adjustbox{raise=0.9cm}{ ,}
\end{equation}
where $l_2=l-\textbf{b}$ and $l_3=l+\textbf{b}$. The choice of using $T_+$ $(L_+)$ or $T_-$ $(L_-)$ is based on whether the vertex opposite to the long side of the triangle is to the right (beginning) or left (end) of the supported link, respectively; this is summarized in the following figure:
\begin{equation}
    \begin{tikzpicture}
        \draw[black, thick] (-2,0)--(2,0);
        \node[isosceles triangle,draw,fill=black, minimum size=0.2cm,inner sep=0pt] at (0,0){};
        \node[above=0.25cm] at (0,0) {$T_-$};
        \node[below=0.25cm] at (0,0) {$T_+$};
        \node[left] at (-2,0) {$L_+$};
        \node[right] at (2,0) {$L_-$};
    \end{tikzpicture}
\end{equation}

For $G=S_3$ the group can be presented as $\braket{r,s|r^3=s^2=e,sr=r^{-1}s}$, giving a 6-dimensional local Hilbert space. $S_3$ has $3$ conjugacy classes $C_e=\{e\}$, $C_r=\{r,r^2\}$, $C_s=\{s,rs,r^2s\}$ with representatives $e$, $r$, and $s$ respectively. The centralizers are $Z(e)\simeq S_3$, $Z(r)\simeq \Z_3$, and $Z(s)\simeq\Z_2$. $\Z_N$ has $N$ irreps (all 1D) and $S_3$ has $2$ 1D irreps and $1$ 2D irrep. Therefore, there are a total of $8$ anyons in the $S_3$ quantum double model. \cite{lo2025universalquantumcomputations3,chen2025universalcircuitsetusing}.
We consider here the maximal decoherence of abelian anyon $([-],C_e)$, where $[-]$ is the non-trivial 1D irrep (sign irrep) of $S_3$. This noise is modeled by the local channel:
\begin{equation}
    \mathcal{N}_l(\rho)=\frac{1}{2}\left( \rho+Q_l\rho Q_l^\dagger \right),
\end{equation}
where the action of $Q$ is\footnote{since $c'\in C_e=\{e\}$, the action on the $l-\textbf{b}$ link is trivial; $Q$ reduces to a single-link operator (it is proportional to the ribbon operator). In fact, $Q$ has an equivalent form to $Z^3$ in $\Z_6$.}:
\begin{equation}
    Q\ket{g}=(-1)^{q_s} \ket{g},\quad q_s=\begin{cases}
        1&g\in\{s,rs,r^2s\},\\0&g\in\{e,r,r^2\}.
    \end{cases}
\end{equation}
The decohered anyon is a boson and has trivial mutual statistics with all other anyons except for $([+],C_s)$ and $([-],C_s)$, where $[+]$ and $[-]$ are the trivial and sign representations of $\Z_2$ respectively \cite{Beigi_2011}, so the resulting theory consists of $6$ types of anyons. The fact that $Q$ is diagonal in the regular representation basis leads to the methodology of sec. \ref{sec:General Protocol} being readily applicable:   
Extend the definition of the on-site tensor $O$ to include a projector for the value of $g$: 
\begin{equation}
O^g_{\alpha\beta\gamma\delta}\rightarrow O^g_{\alpha\beta\gamma\delta\kappa}=O^g_{\alpha\beta\gamma\delta}\delta_{g,\kappa}
\end{equation} 
and connecting tensor $h$ into the TN:
\begin{equation}
    h_{\kappa\kappa'}=\begin{cases}
        1&\kappa\kappa'\in \{e,r,r^2\},\\0&\kappa\kappa'\in \{s,rs,r^2s\},
    \end{cases}
\end{equation}
leading to the following tensor network representation of the maximally-decohered state:
\begin{equation}
    \rho_Q=\sum_{\{g,g'\}}\left[ \prod_lO^{g_l}O^{g'_{l}}h^l \right] \ket{\{g\}}\bra{\{g'\}}.
\end{equation}
This procedure is similar for other channels which decohere abelian anyons, i.e. other choices of $C$ and $\textbf{R}$ such that $|\textbf{R}|=1$ in (\ref{eq:ribbon-h},\ref{eq:ribbon-v}). The procedure is also applicable for decohering non-abelian anyons with trivial mutual statistics with all anyons in the Choi state. For example, decoherence of $([2],C_e)$ (with $[2]$ the 2D irrep of $S_3$) is compatible with this formalism because the anyon has trivial flux; the action is only on a single link, so one can determine $F$ from (\ref{eq:ribbon-h},\ref{eq:ribbon-v}) and construct the connector $h$ all the same.

\subsection{Generalization to arbitrary CSS codes} \label{sec:CSS}

Our construction of fixed-point tensor networks also holds beyond two-dimensional decohered topological phases. In this section, we will discuss the decoherence of arbitrary CSS (Calderbank-Shor-Steane) codes in the magnetic sector.  For our purposes, these are stabilizer codes where all stabilizers $\{A_k\}$, $\{B_j\}$ are tensors of only $\Z_2$ Pauli operators $Z$, $X$, respectively. A TN for the channel
\begin{equation}
    \rho_A \propto \prod_k \frac{1 + A_k}{2}
\end{equation}
is constructed by defining for each stabilizer $A_k$ a tensor $T_{\alpha_1, \dots, \alpha_{n(k)}}^k$, where $\{\alpha_1, \dots, \alpha_{n(k)}\}$ denote the qubits on which the stabilizer $A_k$ is supported. Each index takes values in $\{0,1\}$, and the value of the tensor is
\begin{equation}
    T^k_{\alpha_1, \dots, \alpha_{n(k)}} = \begin{dcases}
        1, & \sum_{i=1}^{n(k)}\alpha_i = 0 \mod 2 \\
        0, & \sum_{i=1}^{n(k)}\alpha_i = 1 \mod 2
    \end{dcases},
\end{equation}
which locally ensures that each term in the bra and ket layer of $\rho_A$ is a simultaneous $+1$ eigenstate of all $A_k$. 
For each qubit $\ell$, we  further define a corresponding tensor $g_{a_1, \dots, a_{m(\ell)}, \omega}^{j_\ell}$, where $\left\{a_1, \dots, a_{m(\ell)}\right\}$ label the stabilizers that act nontrivially on $\ell$, and $\omega$ is the ancillary index used to connect the bra and ket layers. The value of each $g$ tensor is
\begin{equation}
    g^0_{0, \dots, 0, \omega = 0} = g^{1}_{1, \dots, 1, \omega = 1} = 1, \text{ else }0,
\end{equation}
and they are contracted with those $T$ tensors corresponding to the stabilizers labeled by $a_1, \dots, a_{m(\ell)}$. By analogy with earlier examples, we also define a trivial connecting tensor $h^\ell_{\omega \omega'} = \delta_{\omega \omega'}$. The density matrix is expressed after contraction of the virtual indices as
\begin{align}
    \rho \propto & \sum_{j, j'} \left[ \prod_k T^k\left(T^k\right)^* \prod_{\ell}  g^{j_\ell} \left( g^{j'_\ell}\right)^* h^\ell \right] \notag \\
    & \times \ket{j_1, \ldots, j_N}\bra{j_1', \ldots j_N'},
    \label{eq:css-contraction}
\end{align}
which is similar to \eqref{eq:flux_Z2} with trivial $h$ tensor.
A TN for the channel
\begin{equation}
    \rho_B \propto \prod_j \frac{1+B_j}{2}
\end{equation}
can be constructed analogously, in the appropriate basis.

Although arbitrary CSS codes do not necessarily admit an  obvious RG procedure, certain geometries allow for a more straightforward construction. We begin with some general observations. First, let $T_{\alpha_1 \ldots \alpha_n}$ be a tensor of rank $n$ that enforces the $\Z_2$ Gauss law in its virtual indices, i.e.
\begin{equation}\label{eq:gauss-law-general}
    T_{\alpha_1 \ldots \alpha_{n}} = \begin{dcases}
        1, & \sum_{i=1}^{n}\alpha_i = 0 \mod 2 \\
        0 & \sum_{i=1}^{n}\alpha_i = 1 \mod 2
    \end{dcases},
\end{equation}
then for $m < n$ we have that $T$ can be split into the product
\begin{equation}
    T_{\alpha_1 \ldots \alpha_n} = \sum_{\kappa = 0}^1 T_{\alpha_1 \ldots \alpha_m \kappa}^{(1)}T_{\alpha_{m+1} \ldots \alpha_n \kappa}^{(2)},
\end{equation}
where in the $\Z_2$ case we may write
\begin{equation}
    T_{\alpha_1 \ldots \alpha_m \kappa}^{(1)}  = \begin{dcases}
        1, & \kappa + \sum_{i=1}^{m}\alpha_i = 0 \mod 2 \\
        0, & \text{ else}
    \end{dcases},
\end{equation}
and $T^{(2)}$ is defined similarly. Thus we see that the two tensors in the product also symmetrically enforce the $\Z_2$ Gauss law in their indices. Second, let $A$ be a TN with each tensor of the form (\ref{eq:gauss-law-general}), and take a connected subgraph $B \subset A$; clearly, contracting over all indices internal to $B$ yields a new tensor $T_B$ which is proportional to another tensor of the form (\ref{eq:gauss-law-general}).  
Thus since our general recipe for the $\rho_A$ channel in a CSS code produces a TN with all tensors of this form, any RG procedure which only involves splitting and contraction of the form discussed above, and yields the same local geometry, admits the original TN as a fixed point. Below we present the example of three-dimensional $\Z_2$ toric code decohered in the magnetic sector (the $\rho_A$ case).

\subsection{3D toric code renormalization group procedure} \label{sec:3D}

\begin{figure*}[ht]
    \centering
    \includegraphics[width=0.9\textwidth]{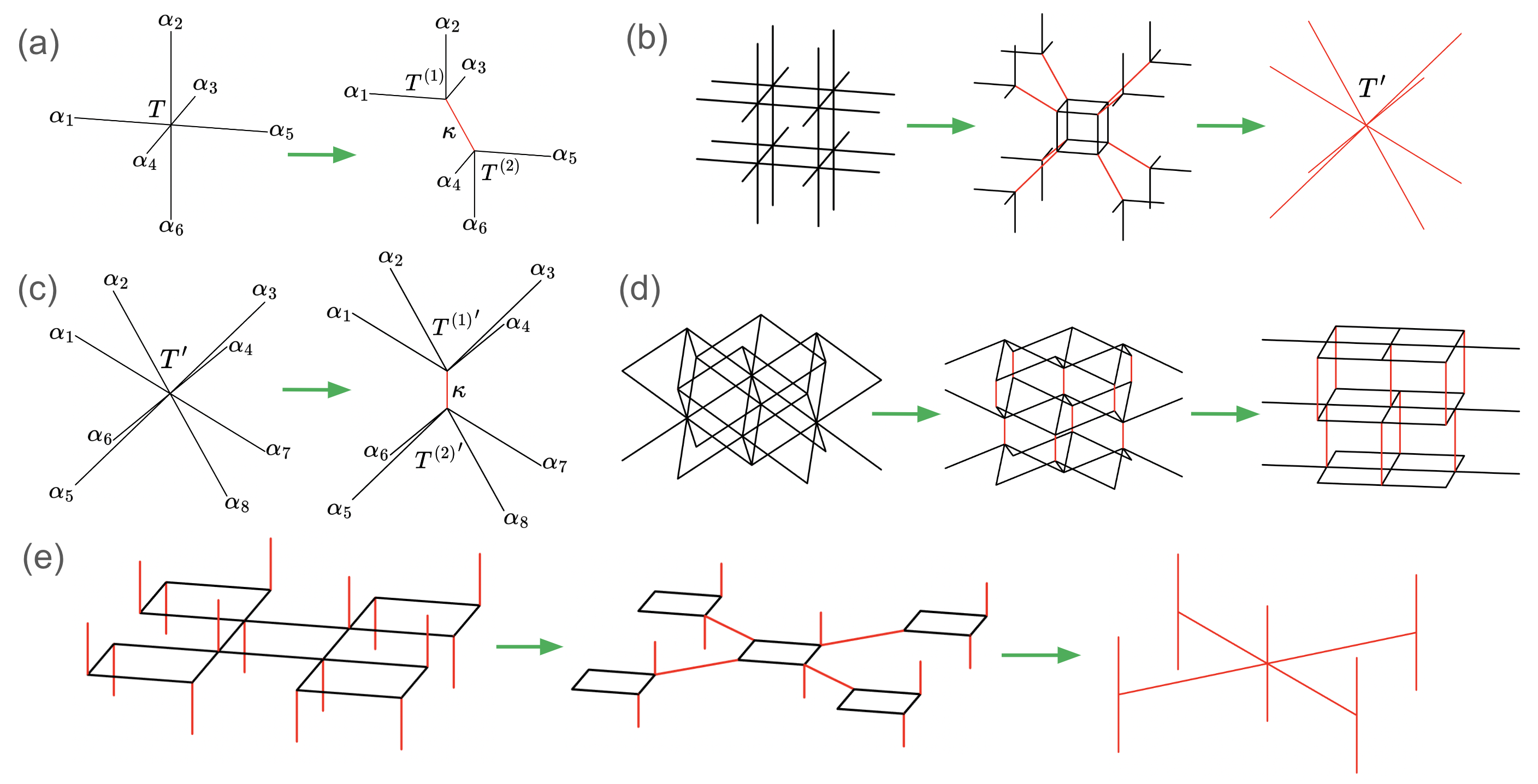}
    \captionsetup{justification=raggedright, singlelinecheck=false}
    \caption{\footnotesize{Multi-step RG procedure for a single (ket) layer of the 3D  toric code TN. (a) The $T$ tensor is split diagonally, in different directions depending on the choice of sublattice. (b) After splitting the TN has the geometry of alternating cubes connected by their vertices. These cubes are contracted over, yielding a new local tensor $T'$. (c) Each $T'$ tensor is split vertically. (d) After splitting and deforming, the TN takes on the geometry of multiple 2D square lattices connected at alternating vertices. (e) Within each 2D layer, RG is performed as in Fig. \ref{fig:RGprocedure}, with the splitting done such that each square has one connection to the upper and lower layer. After contraction, the final TN has the original cubic lattice geometry up to a rotation about the vertical axis.}}
    \label{fig:3d-toric-rg}
\end{figure*}

As a concrete example, we present a RG procedure for the decohered toric code in three spatial dimensions (3D) \cite{He2017EntanglementEF}. The Hamiltonian is defined on a cubic lattice and is analogous to the 2D toric code, with weight-6 $A_v$ stabilizers on vertices and weight-4 $B_p$ stabilizers on square faces; the corresponding TN hence has rank 6 tensors $T_{\alpha_1, \ldots, \alpha_6}$ on each vertex, and tensors $g_{a_1, a_2, \omega}^{j_\ell}$ on each edge $\ell$. The RG procedure here is less geometrically straightforward than the 2D case, involving multiple distinct steps, and is illustrated in Fig.\ \ref{fig:3d-toric-rg}.

First, at each vertex in the bra and ket layer, the local tensor $T$ is split into the product
\begin{equation}
    T_{\alpha_1 \ldots \alpha_6} = \sum_\kappa T_{\alpha_1 \alpha_2 \alpha_3 \kappa}^{(1)}T_{\alpha_4 \alpha_5 \alpha_6 \kappa}^{(2)},
\end{equation}
with the direction of the splitting chosen such that geometrically the new TN resembles an alternating lattice of cubes, connected at their corners. Then, as before, the two layers of the mixed state TN are contracted over the physical indices $j_\ell$ of the $g$ tensors, and we contract over the virtual indices in each cube, to form a new TN with a different geometry, and doubled tensors $\widetilde{T}_{\alpha_1 \alpha_1' \ldots \alpha_8 \alpha_8'}'$. It is again easily checked that $\widetilde{T}'$ is proportional to a product of two copies of the rank 8 tensor $T'_{\alpha_1 \ldots \alpha_8}$; hence for clarity, from here we work only in a single (ket) layer. In the second stage, each $T'$ is split vertically into the product
\begin{equation}
    T_{\alpha_1 \ldots \alpha_8}' = \sum_\kappa T_{\alpha_1 \ldots \alpha_4 \kappa}^{(1)'}T_{\alpha_5 \ldots \alpha_8 \kappa}^{(2)'},
\end{equation}
and as shown in Fig.\ \ref{fig:3d-toric-rg}, the resulting TN now has the geometry of multiple layers of 2D square lattices, connected via alternating vertices. From here, we proceed within each layer as in the 2D case, splitting such that each square has a single link to the upper and lower layer. Contracting over these squares then yields the original cubic lattice geometry. 
Note that since each of the original $T$ tensors enforces the $\Z_2$ Gauss law in its 6 indices, and each of the splitting and contraction operations here again yield Gauss law tensors with different ranks, the original TN is a fixed point of the RG procedure.

\subsection{Tensor network representation of mixed-state chiral semion through decoherence}\label{sec:chiral}

In this part, we provide a fixed-point tensor network representation of the chiral semion theory. While such a topological order is allowed in pure states,  it is believed to not arise in fixed-point wavefunctions for locally decomposable Hilbert spaces $\mathcal{H}=\otimes_l \mathcal{H}_l$ of finite dimension (see for example \cite{Kitaev_2006,PRXQuantum.6.010313}). Therefore, the fixed-point tensor network is unique in the mixed-state setting. 

We start from the pure-state double semion tensor network representation, which was derived in section \ref{sec:DS pure} from condensing $e^2m^2$ in $\Z_4$ toric code, and then decohere in the $s=em$ channel.
The initial pure state in the doubled space is (see eq. \ref{eq:psi_cond}):
\begin{align}
   |\rho_0\rangle\!\rangle = \sum_{\{x,z,x',z'\}}& \left[ \prod_v\widetilde{T}^v\prod_lg^{(x_l,z_l)} g^{(x'_{l},z'_{l})}\tilde{h}^l \right]\notag\\&\times\ket{\{ (x,z)\} }\otimes\ket{\{ (x',z')\}},
\end{align}
with $\tilde{h}^l=h^lh^{*l}$. Let $\widetilde{U}$ denote here the basis transformation from $\widetilde{X},\widetilde{Z}$ to $X^2,Z^2,\overbar{X}^2,\overbar{Z}^2$. The new basis will be denoted $(a,b)$, so $\widetilde{U}^{(a,b)}_{x,z,x',z'}=\braket{a,b|x,z,x',z'}$. Then ,as before, we combine the two single-link tensors on the bra and ket layers:
\begin{equation}
\widetilde{g}^{(a,b)}_{\alpha\beta\alpha'\beta',\mu\nu\mu'\nu'}=\sum_{x,z,x',z'}g^{(x,z)}_{\alpha\beta,\mu\nu}g^{(x',z')}_{\alpha'\beta',\mu'\nu'} \widetilde{U}^{(a,b)}_{x,z,x',z'}.
\end{equation}
The initial state can now be written as:
\begin{equation}
    |\rho_0\rangle\!\rangle =\sum_{\{a,b\}}\left[ \prod_v\widetilde{T}_{v}\prod_l\widetilde{g}^{(a_l,b_l)} \prod_l \widetilde{h}^l\right]\ket{\{a,b\}}
\end{equation}

The local decoherence channel is:
\begin{equation}
    \mathcal{N}_l\left[(\cdot)\right]=\frac{1}{4}\sum_{j=0}^3X^j_lZ^j_{l-\textbf{a}}(\cdot) {X^\dagger}^j_l{Z^\dagger}^j_{l-\textbf{a}},
\end{equation}
which is of the required projector form mentioned in (\ref{eq: general choi projector}) when considering the doubled space. Since the initial density matrix is stabilized by $X_l^2 Z_{l-\bm{a}}^2,$ the channel reduces to:
\begin{equation}
    \mathcal{N}_l\left[(\cdot)\right]=\frac{1}{2}\left( (\cdot)+X_lZ_{l-\textbf{a}}(\cdot) X^\dagger_lZ^\dagger_{l-\textbf{a}} \right).
\end{equation}

To take into account the decoherence, extend the definition of $\widetilde{g}$ with projector ancilla as before, $\widetilde{g}^{(a,b)}_{...}\rightarrow\widetilde{g}^{(a,b)}_{...,\varepsilon\omega}=\delta_{a,\varepsilon}\delta_{b,\omega}\widetilde{g}^{(a,b)}_{...}$, and introduce the connecting tensor to enforce the projection from the channel:
\begin{equation}
    \mathfrak{h}_{\varepsilon_l\omega_{l-\textbf{a}}}=\begin{cases}
        1&(\varepsilon_l+\omega_{l-\textbf{a}})\text{ mod }4=0,\\0&(\varepsilon_l+\omega_{l-\textbf{a}})\text{ mod }4\neq0,
    \end{cases}
\end{equation}
i.e. the TN consists of two types of connectors, $\tilde{h}$ (for the condensation in the pure state) and $\mathfrak{h}$ (for the decoherence in the mixed state). The TN representation for the post-decoherence Choi state is then:
\begin{align}
     |\rho_\text{CS}\rangle\!\rangle\propto& \sum_{\{a,b\}} \left[ \prod_v\widetilde{T}^v\prod_l\widetilde{g}^{(a_l,b_l)}\tilde{h}^l\mathfrak{h}^l \right]\notag\\&\times\ket{\{(a,b)\}}.
\end{align}
The physical density matrix can then be obtained as in \eqref{eq: physical density matrix}. It is chiral from the perspective of anyon contents and symmetry; in particular, the chiral semion 1-form symmetry is spontaneously broken from strong to trivial, while the antichiral semion sector is weakly symmetric, in the sense of \cite{zhang2025strongtoweakspontaneousbreaking1form}.

\section{Beyond Fixed-Points}\label{sec:Z3 numerics}

In this section, we extend the tensor network formalism beyond the fixed point limit, considering incoherent error channels of non-maximal strength. To illustrate the idea, we focus on the $m^ae^b$ decoherence of $\mathbb{Z}_N$ toric code, but the extension applies to all examples discussed in this work. At fixed point, the connecting tensor in \eqref{eq:most general h} simply projects to $a\mu_l+b\nu_{l-\textbf{a}}=0$ mod $N$, arising from the fact that maximal decoherence is equivalent to projective measurements of the Kraus operators (after summing over measurement outcomes). Turning to non-maximal decoherence strengths intuitively corresponds to weakening the projection. More concretely, choose the simplest local error channel creating only pairs of $m^ae^b$, 
\begin{equation}
    \mathcal{N}_l[\rho]=(1-p)\rho+pX^a_lZ^b_{l-\mathbf{a}}\rho X^{-a}_lZ^{-b}_{l-\mathbf{a}}.
 \label{eq:non_fixed}
\end{equation}
The corresponding nontrivial doubled Kraus operator is $\widetilde{K}_{1,l}= p X^a_lZ^b_{l-\mathbf{a}} \overline{X^a_l} \overline{Z^b_{l-\mathbf{a}}}$, supported locally near link $l$. By construction, the eigenstates of $\widetilde{K}_{1,l}$ are $\ket{\{x_l,z_l\}}$, and the  eigenvalues are given by $k_{1,l}=p\ \omega^{a x_l +b z_l}$. Similarly, the eigenvalue of the trivial doubled Kraus operator is simply $k_{0,l}=1-p$.  The decohered Choi state can thus be written in this basis as
\begin{equation}
 \ket{\rho}\!\rangle \propto\sum_{\{x_l,z_l\}} \left[ \prod_v\widetilde{T}^v \prod_l\widetilde{g}^{(x_l,z_l)} \right] \prod_l \sum_{\alpha=0,1} k_{\alpha ,l} \ket{\{x_l,z_l\}},
\end{equation}
where the summation over $\alpha$ gives $(1-p)+p\ \omega^{a x_l + b z_l} $. 
Next, following the steps in the tensor network construction in Fig. \ref{fig:flowchart} to extend $\tilde{g}$ by adding ancillas, we arrive at the connecting tensor 
\begin{equation}
 h_{\mu_l\nu_{l-\textbf{a}}}=(1-p)+p\ \omega^{a \mu_l + \nu_{l-\mathbf{a}}}. 
\end{equation} 
Then one can complete the remaining steps: perform the local inverse basis change, return to the physical Hilbert space and arrive at the tensor network representation of the mixed state itself. To connect  with the previous results, take for example $N=2$ and $a=b=1$, in the strong decoherence limit $p=1/2$ the connecting tensor reduces to the previously studied case of fixed-point fermion decoherence channel where $ h_{\mu_l\nu_{l-\textbf{a}}}=\frac{1}{2} (1-\mu_l -\nu_{l-\textbf{a}})$ projects out the configurations which satisfy $\mu_l + \nu_{l-\textbf{a}}=1$ $\mod 2$.  If more types of anyons are created by the incoherent channel, one can simply add more terms to \eqref{eq:non_fixed}.

\subsection{Particle-Hole Symmetric Channel} \label{sec:Z3 TEN and QRE}

Below we consider a concrete example where the extended formalism is employed for particle-hole asymmetric decoherence of the $\mathbb{Z}_3$ toric code. This case, to the best of our knowledge, has not been examined in existing literature, partially due to the sign problem. The local error channel reads 
\begin{equation}
    \mathcal{N}^{p_1,p_2}_\ell[\rho]=p_0\rho + p_1 X_\ell\rho X_\ell^2 + p_2X_\ell^2 \rho X_\ell, \label{eq:Z3 X}
\end{equation}
where $p_0=1-p_1-p_2$. 

The decoherence transition separates a phase where the $\mathbb{Z}_3$ toric code contains $9$ types of Abelian anyons, from a post-decohered state which hosts 3 types of transparent Abelian anyons and  no topological quantum memory.  Channel (\ref{eq:Z3 X}) represents pure flux decoherence, and so its TN representation is the $\mathbb{Z}_3$ analogue of (\ref{eq:flux_Z2}), with $h^l=(1-p_1-p_2) + p_1 j_l j'^2_l + p_2j^2_lj'_l$.

The decoherence can be diagnosed via entanglement measures such as the topological entanglement negativity \cite{PRXQuantum.5.020343,PhysRevA.65.032314,PhysRevA.88.042319} and the quantum relative entropy (QRE) \cite{10.2996/kmj/1138844604,PRXQuantum.5.020343}.For example, along the symmetric line $p_1=p_2$, these quantities can be mapped to coupled replicas of 3-state Potts models, and the transition can be diagnosed via Monte Carlo. In contrast, off the symmetric line $(p_1\neq p_2)$, the mapping results in a non-Hermitian Hamiltonian with a sign problem \cite{PRXQuantum.5.020343}. While unbiased Monte Carlo is no longer available, the TN method still allows us to determine the R\'enyi transition. We compute the QRE given by:
\begin{equation}\label{eq:QRE}
    D^{(n)}(\rho||\rho_\alpha)=\frac{1}{1-n}\log\frac{\text{Tr}(\rho\rho_\alpha^{n-1})}{\text{Tr}(\rho^n)}.
\end{equation}
where $\rho=\mathcal{N}[\rho_0]$ and $\rho_\alpha=\mathcal{N}[w_\alpha(P)\rho_0w_\alpha^\dagger(P)]$, and $w_\alpha(P)$ creates an anyon $\alpha$ and its antiparticle $\bar{\alpha}$ at the endpoints of path $P$. The R\'enyi-4 QRE essentially measures the post-decoherence distinguishability of a state free of excitations from a state in the presence of excitations. Below the critical error threshold, the QRE diverges in the thermodynamic limit as the distance between anyons (the length of the path $|P|$) increases. Above the threshold, it would saturate to a finite value with growing distance \cite{PRXQuantum.5.020343}. In order to probe distinguishability, the anyons in (\ref{eq:QRE}) are chosen to match those that remain after decoherence; the channel (\ref{eq:Z3 X}) with maximal decoherence results in an anyon theory with $\{1,m\}$, so $\alpha$ is taken to be $m$.

To facilitate the calculation, we place the system on a cylinder of circumference $L$ and apply the transfer matrix formalism, representing the contraction of the TN as consecutive transfer matrix products along the infinite direction (a single transfer matrix is simply one row of the TN along the circumference of the cylinder). We use the TN method to calculate the QRE for different decoherence strengths $0.05\leq p_2\leq0.475$ at fixed $p_1=0.05$; results for $p_2$ outside this interval are related to these via products of on-site unitary operators which preserve QRE. 
Let $T_0$ and $T_\alpha$ represent the transfer matrix without and with the presence of the excitation string respectively. The TN contraction representing $\text{Tr}(\rho\rho_\alpha^{n-1})$ and $\text{Tr}(\rho^n)$ are both representable by repeated transfer matrix products of $T_0$ and $T_\alpha$, and they only differ along the line where the excitation is present in the numerator. Orient the path $P$ along the infinite direction (no segments of $P$ are parallel to the periodic direction). For large enough $|P|$, the behavior of these repeated products of transfer matrices is controlled by their largest eigenvalues $\lambda_0$ and $\lambda_\alpha$, given by:
\begin{equation}
T_0r_0=\lambda_0r_0,\;\;l^T_0T_0=\lambda_0l^T_0,
\end{equation}
where $l_0,r_0$ are the left and right eigenvectors of $T_0$, and the same can be defined for $T_\alpha$. These eigenvalues are positive and unique since both $T_0$ and $T_\alpha$ satisfy the Perron-Frobenius conditions \footnote{Upon re-expressing the contracted network with error configuration variables akin to the method in Appendix \ref{sec:Tr rho n}, the matrix elements of the transfer matrices can be made strictly positive.} The factors in the numerator and denominator of (\ref{eq:QRE}) cancel away from the anyonic string, and so:
\begin{equation}
    D^{(n)}(\rho||\rho_\alpha)\approx -\frac{1}{n-1}\log\frac{l_0^TT^{|P|}_\alpha r_0}{l_0^TT^{|P|}_0r_0}.
\end{equation}
For large $|P|$, we can substitute $T_\alpha^{|P|}\approx \lambda_\alpha^{|P|} \frac{r_\alpha l_\alpha^T}{l_\alpha^Tr_\alpha}$, leading to:
\begin{equation}
    D^{(n)}(\rho||\rho_\alpha)\approx\frac{\sigma_L}{n-1} |P| -\frac{1}{n-1}\log C_\alpha +o(1),
\end{equation}
where $\sigma_L=\log \frac{\lambda_0}{\lambda_\alpha}$ and $C_\alpha$ is a constant with respect to $|P|$, dependent on the form of the dominant eigenvectors of $T_0$ and $T_\alpha$. Since the ratio of the traces is controlled by $(\lambda_0/\lambda_\alpha)^{|P|}\approx e^{|P|\sigma_L}$, scale invariance at the critical point implies that $\sigma_L$ scales as $1/L$. Therefore, we plot $L\sigma_L$ as a function of $p_2$ and locate the mutual intersection point for multiple $L$ to minimize finite-size effects.

\begin{figure}
    \centering
    \includegraphics[width=0.95\columnwidth]{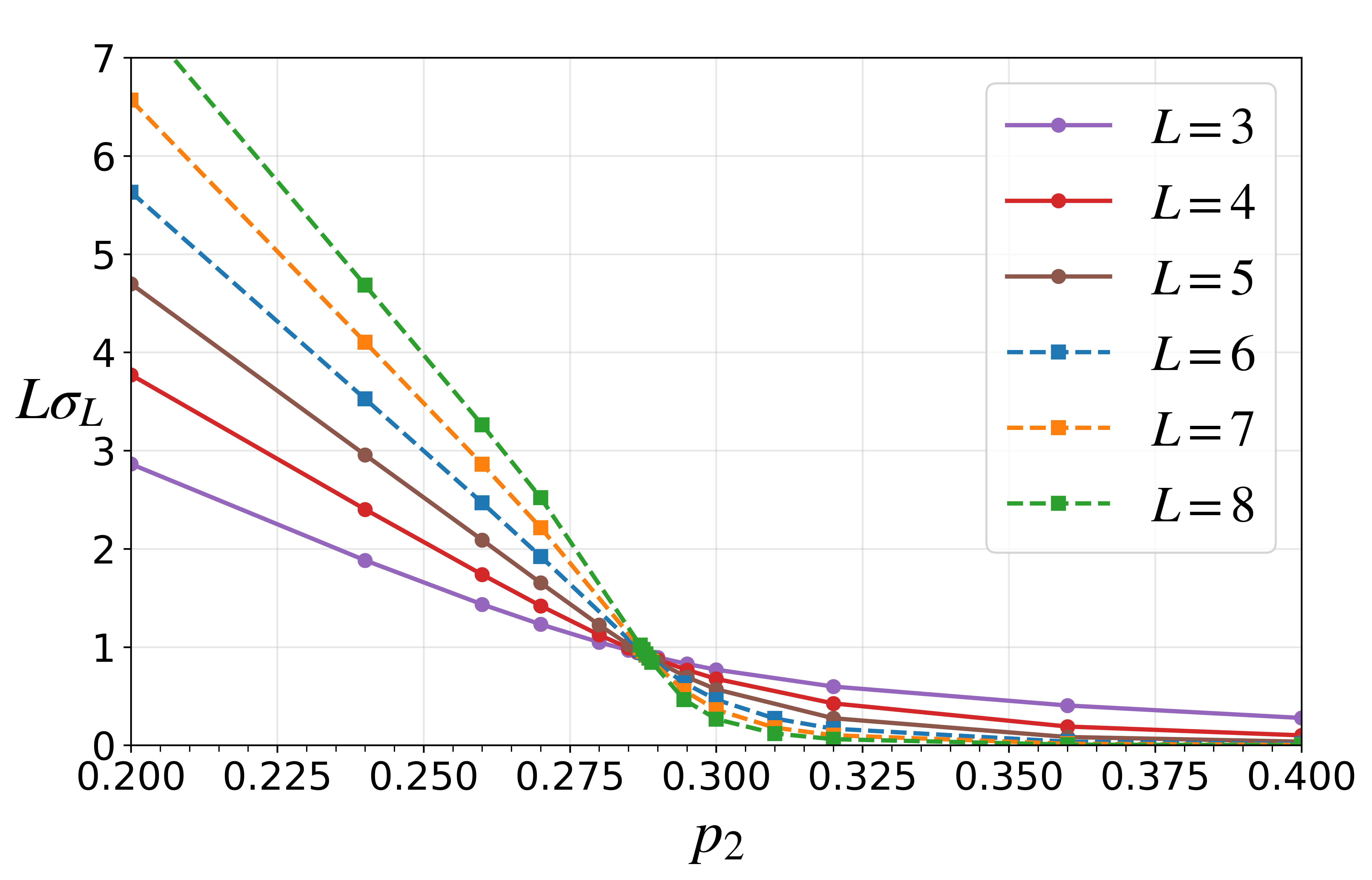}
    \captionsetup{justification=raggedright, singlelinecheck=false}
    \caption{\footnotesize{Quantum relative entropy calculation: $L\sigma_L$ for $n=4$ as a function of $p_2$ for various system sizes. The tranfer matrix can be diagonalized for system sizes $L \leq 5$ (solid lines), beyond that requiring bond dimension truncations (dashed lines). Bond dimension limit $\chi=128$ is used for $L>5$. The transition point is $p_2\approx0.288\pm0.001$.}}
    \label{fig:LsigmaL}
\end{figure}
The QRE results for $n=4$ are shown in Fig. \ref{fig:LsigmaL}, and the transition is estimated to be $p_2\approx0.288$. The effect of bond dimension truncation for larger $L$ is roughly on par with finite size effects, and more details are presented in Appendix \ref{sec:Z3 numerics supplement}. The Appendix also addresses the corresponding topological entanglement negativity calculation, which give consistent results. One can perform the same analysis for various other R\'enyi indices $n$ as well, which we leave for future research.

\section{Discussion and Outlook}
\label{sec:discussion}

We presented a systematic formalism, summarized for fixed-points in Fig. \ref{fig:flowchart}, to construct tensor network representations for decohered topological phases and more general CSS codes at arbitrary decoherence strength for a large class of incoherent noise. 
This framework can be efficiently employed in various scenarios where Monte Carlo simulations become unavailable due to sign problems, such as in the study of critical error thresholds for particle-hole asymmetric noise discussed in section \ref{sec:Z3 numerics}, as well as in the stability study of mixed-state topological phases similar to the pure-state stories in \cite{PhysRevB.104.235151,PhysRevB.95.235119,zhang2020non,PhysRevLett.124.130603,PhysRevLett.130.216704}. 
The scope of the method is readily generalizable beyond the examples presented, for example to string-net models \cite{Levin_2005}. It can also be applied directly to the decoherence of states that are themselves the result of condensation, as given in the example of the chiral theory's tensor network in section \ref{sec:chiral}.

The limitation of the method can be seen as follows: In general, the commutator between different Kraus operators can be described by a matrix $U$ instead of simply a phase factor, such that in the doubled space, $K_1K_2\otimes \overbar{K_1}\overbar{K_2}=UK_2K_1\otimes \overbar{U}\overbar{K_2}\overbar{K_1}$ and $U, \overline{U}$ do not cancel out. Therefore, one cannot find a preferable basis where the doubled Kraus operators are simultaneously diagonalized, such that step (i) in Fig. \ref{fig:flowchart} is not directly possible. Similar problems occur for coherent noise: with the exception of the simple case of ultra-local channels (where each Kraus operator only acts on a single link degree of freedom), such noise is generically not commutative up to a phase.

One striking feature of the construction is the existence of a fixed-point tensor network for the chiral semion mixed state (section \ref{sec:chiral}). In pure states, chiral topological order is believed to be incompatible with exact renormalization-group fixed-point wavefunctions on a lattice with finite-dimensional on-site Hilbert spaces (see \cite{niu2026} for the most recent progress closely related to mixed states), or a gapped local commuting projector parent Hamiltonians \cite{Kitaev_2006,PhysRevB.101.045137}. That this obstruction is absent for the decohered state raises the question of what ``chirality'' means once the state is mixed. Following the recent work of \cite{Chiral_Luo,ellison2026}, it is expected that the mixed state chiral semion can, on a geometry with boundary, be written as a product of local commuting projectors, and that the boundary is only short-range correlated. Whether this state is connected to the pure-state chiral semion topological order by a finite-depth local channel remains open, and we regard it as a sharp question partially forced by our results. At the level of the anyon content and its braiding data, \cite{Chiral_Luo} proposed relative-entropy-based order parameters for chirality, and such quantities are accessible from our tensor network representations, similar to the computations presented in section \ref{sec:Z3 numerics}. 

The full boundary data itself is a more delicate matter. For pure-state tensor networks, the virtual indices along a cut carry an MPO algebra whose representation theory encodes the universal boundary data, and the entanglement spectrum inherits the corresponding structure (see for example \cite{Schuch_2013}). In mixed states, 
as discussed for the above example of mixed-state chiral topological orders, a long-range correlated boundary need not accompany a chiral bulk, which is in contrary to the case of the corresponding pure state chiral topological order. Therefore, the boundary data may not be an invariant of the mixed-state phase in the sense of two-way connectivity by finite depth local channels \cite{Coser_2019}. We henceforth regard the fundamental question of bulk-boundary relations in mixed-state topological order as important and conceptually open. Nevertheless, we expect the tensor network representations constructed here to provide a natural and concrete setting in which the universal boundary data can be extracted, whatever that data might be.

Before ending, we comment on the relation between the anyon excitations in the decohered theory and the virtual symmetries of the tensor network. Take our favorite $\mathbb{Z}_N$ toric code example: In pure states \cite{Schuch_2010,Bultinck_2017}, the Gauss-law $T$ tensor around the four-legged vertex obeys
\begin{equation}
\big(Z^{k}\otimes Z^{-k}\otimes Z^{-k}\otimes Z^{k}\big)\,T \;=\; T ,
\quad k\in\mathbb{Z}_N,
\label{eq:virtsym}
\end{equation}
matching the sign pattern of $\alpha-\beta-\gamma+\delta$ in \eqref{eq:T_pure}. Pushing this $\mathbb{Z}_N$ virtual symmetry out through the four adjacent tensors $g$ leads to the shortest closed physical $e$-string encircling that vertex. Deformability of $e$-strings is the pull-through consequence $Z_\alpha Z_\beta^{-1}T = Z_\gamma Z_\delta^{-1}T$ of Eq.~\eqref{eq:virtsym}, such that  contractible closed $e$-loops act as the identity and an open virtual $Z$-string terminates on a pair of $e$'s. By comparison, a virtual $X$-string threading a sequence of $T$'s shifts two legs with opposite signs and is therefore transparent to $T$. However, since $g$ simply copies the virtual index to the physical, the virtual $X$ string acts nontrivially on the physical index after pushing out, and its endpoints are the $m$ excitations. Moving on to mixed states, before decoherence, the doubled tensors $\widetilde{T}=T\otimes T^{*}$ and $\widetilde{g}$ inherit the virtual symmetries in each layer separately. The question of which virtual symmetries survive decoherence is determined by the connecting tensor $h$. 
Suppose the local Kraus operator is $S_l=X_l^{a}Z_{l-\mathbf{a}}^{b}$, such that $h$ implements the projector onto $\widetilde{S}_l=+1$ with $\widetilde{S}_l =S_l\otimes\overline{S}_l$. Let $W_\lambda(\gamma)$ be the closed
virtual string for $\lambda=m^{\alpha}e^{\beta}$ on loop $\gamma$.  Then: 
 If $W_{\lambda}$ commutes with every $\tilde{S}_l$, i.e. $a\beta-b\alpha=0$ $\mod N$, then $\lambda$ remains as deconfined excitations in the decohered theory, corresponding to a strong 1-form symmetry.  Else, $\lambda$ becomes confined and corresponds to a weak 1-form symmetry. 

\begin{acknowledgements}
We thank Junyeong Ahn, Pochung Chen, David P. Garc\'{i}a, Yingfei Gu, Chao-Ming Jian, Itamar Kimchi, Yuhan Liu, Abhinav Prem, Shijun Sun, Xiaoqi Sun, Xueda Wen, Carolyn Zhang, and Jianhao Zhang for helpful conversations. Zhu-Xi thanks Yingfei Gu, Zhicheng Yang and Yu-An Chen for their hospitality during her visits in June 2025 while this work was being finalized. The Sol and Astra models of ChatGPT were used for parts of the code relevant to section \ref{sec:Z3 numerics}, which was consistency-checked intermediately throughout the calculations for validation, and compared with known results for verification as seen in Appendix \ref{sec:Z3 numerics supplement}. 
\end{acknowledgements}

\begin{appendix}

\section{TERG-invariance of \texorpdfstring{$\text{Tr}(\rho^n)$}{Tr(rho power n)}}\label{sec:Tr rho n}

The RG procedure for $\text{Tr}(\rho^n)$ with $n>1$ requires an initial change of basis to an ``error configuration'' basis to allow for a generalization of the TERG protocol reviewed in sec. \ref{sec:1+Av RG}. We begin with the $n=2$ case for simplicity, where $\text{Tr}(\rho^2)=\langle\!\langle\rho|\rho\rangle\!\rangle$. As in sec. \ref{sec:XZ-type channel}, we consider a general dyonic channel of the form (\ref{eq:double projector}).  The connecting tensor $h_{\mu_l \nu_{l-\mathbf{a}}}$ given in (\ref{eq:most general h}) has following decomposition:
\begin{equation} \label{eq:h fourier}
    h_{\mu_l \nu_{l-\mathbf{a}}}=\frac{1}{N}\sum_{q=0}^{N-1} \omega_N^{q(a\mu_l+b\nu_{l-\mathbf{a}})},
\end{equation}
with $\omega_N$ the $N^\text{th}$ root of unity as before. Let the doubled Kraus operator $\widetilde{D}_l=\widetilde{X}^a_l\widetilde{Z}^b_{l-\mathbf{a}}$, where the doubled operator is defined as $\widetilde{O}=O\otimes\bar{O}$ as in sec.\ \ref{sec:General Protocol}. Then using (\ref{eq:h fourier}) the product of all connecting tensors, which represents the channel action, can be expressed as:
\begin{equation}
    \prod_lh_{\mu_l \nu_{l-\mathbf{a}}} = \frac{1}{N^M} \sum_{\{q_l\}} \bra{\{\mu,\nu\}} \prod_l \widetilde{D}_l^{q_l} \ket{\{\mu,\nu\}},
\end{equation}
where $M=2L^2$ 
as usual. Then the trace can be expressed as:
\begin{equation}
    \text{Tr}(\rho^2)\propto \sum_{\{q_l\}}  \langle\!\bra{\rho_0} \prod_l \widetilde{D}^{q_l}_l \ket{\rho_0}\!\rangle.
\end{equation}
For each configuration $\{q\}$, the term will vanish unless $\prod_l \widetilde{D}^{q_l}_l$ does not create open strings of anyons. A local error configuration $q_l$ can be associated to the midpoint of the blue string connecting the links $(l,l-\mathbf{a})$ in the image below: 

\begin{equation}
    \begin{tikzpicture}
        \draw[Black,thick] (-2,0)--(2,0);
        \draw[Black,thick] (0,-2)--(0,2);
        \draw[Black,thick] (2,0)--(2,2);
        \draw[Black,thick] (0,2)--(2,2);
        \draw[Blue,thick] (1,2)--(-1,0);
        \draw[Blue,thick] (2,1)--(0,-1);
        \node[above left] at (0.575,1.5) {\small{\textcolor{Blue}{$q_\text{U}$}}};
        \node[below right] at (1.45,0.55) {\small{\textcolor{Blue}{$q_\text{R}$}}};
        \node[above left] at (-0.5,0.5) {\small{\textcolor{Blue}{$q_\text{L}$}}};
        \node[below right] at (0.5,-0.5) {\small{\textcolor{Blue}{$q_\text{D}$}}};
        \node[below left] at (0.05,0.05) {\small{$v$}};
        \node[] at (0.5,0.5) {\small{\textcolor{Blue}{$t$}}};
        \draw[Blue,dashed] (-0.5,1.5)--(1.5,1.5)--(1.5,-0.5)--(-0.5,-0.5)--cycle;
        \node[circle,draw,fill=Blue, minimum size=3pt,inner sep=0pt] at (0.5,1.5){};
        \node[circle,draw,fill=Blue, minimum size=3pt,inner sep=0pt] at (1.5,0.5){};
        \node[circle,draw,fill=Blue, minimum size=3pt,inner sep=0pt] at (0.5,-0.5){};
        \node[circle,draw,fill=Blue, minimum size=3pt,inner sep=0pt] at (-0.5,0.5){};
    \end{tikzpicture}
\end{equation}
for there to be no excitations, a Gauss law must be enforced at the level of the error configurations, which requires shifting the original vertex tensor $T$ to a tensor $t$ at the center of the vertex-plaquette pair $(v,p(v))$, where we maintain the convention of oriented strings entering the tensor from the left and below and exiting from above and the right. Let $s=q_\text{L}+q_\text{D}-q_\text{R}-q_\text{U}$. For the newly defined tensor to impose the constraint required by equation (A3), it must satisfy $as \text{ mod } N=bs \text{ mod } N=0$. If we define $d=\frac{N}{\text{gcd}(a,b,N)}$, then the condition reduces to $s \text{ mod } d=0$. Therefore, the shifted tensor network is described by:
\begin{equation}
    t_{\alpha\beta\gamma\delta} = \begin{cases} 1 & \alpha-\beta-\gamma+\delta \; \: \text{mod }d=0,  \\ 0 & \text{else}, \end{cases}
\end{equation}
and $f^q_{\alpha\beta}=\delta^q_{\alpha\beta}$. In this new basis and shifted tensor network, the TERG procedure described in sec. \ref{sec:1+Av RG} proceeds as before.

The above can be generalized to arbitrary $n$. In this case for replica copy $r$, the density matrix reads:
\begin{equation}
    \rho_{\{q^{(r)}\}}=\frac{1}{N^M} \sum_{\{q^{(r)}\}} D(\{q^{(r)}\}) \rho_0 D^\dagger(\{q^{(r)}\}),
\end{equation}
Then:
\begin{equation}
    \text{Tr}(\rho^n)= \!\!\!\!\! \sum_{\{q^{(1)},q^{(2)},\dots, q^{(n)}\}} \!\!\!\!\!\! \text{Tr}\bigg[ \rho_{\{q^{(1)}\}}\rho_{\{q^{(2)}\}}\cdots\rho_{\{q^{(n)}\}} \bigg].
\end{equation}
The non-vanishing condition of the trace now requires a Gauss law of the \textit{relative} errors $k^{(r)}\equiv q^{(r)}-q^{(r+1)}$. Thus, the tensor network is described by the replicated and shifted $f^{(r)}$ and $t^{(r)}$ tensors:
\begin{equation}
    \text{Tr}(\rho^n)\propto \sum_{\{k^{(1)},\dots, k^{(n-1)}\}} \prod_{r=1}^{n-1}\bigg[ \prod_vt_v^{(r)} \prod_l f^{k_l^{(r)}}\bigg],
\end{equation}
where the RG will again proceed from here as outlined in sec. \ref{sec:1+Av RG}.

\section{Fermionic tensor network approach} \label{sec:bosonization}

We mention here an alternative tensor network formulation for the the $\Z_2$ toric code decohered in the fermion channel, which makes use of the fermionic projected entangled pair state (fPEPS) construction \cite{Kraus_2010}. Consider a system defined on a square lattice with fermions living on the plaquettes. There is an isomorphism between the even-parity subspace of this system and the subspace of the original system where $A_vB_{p(v)}$ acts as the identity \cite{Chen_2018}, where $p(v)$ is the plaquette to the top-right of vertex $v$. The isomorphism is given by the following map:
\begin{align}
    B_p\longleftrightarrow (-1)^{f^\dagger_pf_p}=-im_pm'_p,\notag\\
    X_lZ_{l-\textbf{a}}\longleftrightarrow im_{L(l)}m'_{R(l)},
\end{align}
where $l-\textbf{a}$ is again the neighboring link below and to the left of link $l$, $f_p$ annihilates a complex fermion at plaquette $p$, $m_p=f_p+f^\dagger_p$ and $m'_p=-i(f_p-f^\dagger_p)$ are the Majorana fermions corresponding to $f_p$, and $L(l)$, $R(l)$ are the plaquettes to the left and right of link $l$ relative to its orientation. Then the maximally decohered state can be expressed as a sum over the even-parity states \cite{PRXQuantum.6.010315}:
\begin{equation} 
    \rho=\frac{1}{2^{N_p-1}} \sum_{\{n_p\} }^\text{even{}} \ket{\{n_p\}} \bra{\{n_p\}},
\end{equation}
with $n_p=f^\dagger_pf_p$, and $N_p$ is the number of plaquettes. A \textit{fermionic} tensor network description of the above state can be given as follows \cite{Kraus_2010}: The virtual spaces (i.e. the Hilbert spaces for the degrees of freedom that are to be contracted over) are taken to be fermionic here, so define one virtual fermionic mode $(a^\dagger,b^\dagger,c^\dagger,d^\dagger)$ for each leg of tensor $T^k_{\alpha\beta\gamma\delta}$, which are now defined on plaquettes instead of vertices, and carry a physical leg $k$ for the physical fermion $f$. Now define  maximally entangled creation operators on plaquettes separated horizontally ($H$) and vertically ($V$), and projection $(Q)$ operators on each plaquette:
\begin{align}
    H_{(i,j)}=\frac{1}{\sqrt{2}}(1+c^\dagger_{(i,j)}a^\dagger_{(i+1,j)}),\notag\\V_{(i,j)}=\frac{1}{\sqrt{2}}(1+b^\dagger_{(i,j)}d^\dagger_{(i,j+1)}),\notag\\
    Q_p= \sum(T_p)^k_{\alpha\beta\gamma\beta} f^k_p a^\alpha_p b^\beta_p c^\gamma_p d^\beta_p,
\end{align}
where $(i,j)$ label the $(x,y)$ coordinates of the plaquettes (the directional conventions here match those in Fig. \ref{fig:Ttensor}), and $T$ is given by:
\begin{equation}
    T_{\alpha\beta\gamma\delta}^k=\begin{cases}1&(k+\alpha-\beta-\gamma+\delta)\text{ mod }2=0,\\0&(k+\alpha-\beta-\gamma+\delta)\text{ mod }2\neq0.\end{cases}
\end{equation}
This restricts to the even-parity sector. A pure state representing $\frac{1}{2^{N_p-1}} \sum_{\{n_p\} }^\text{even{}} \ket{\{n_p\}}$ can then be expressed as:
\begin{equation}
    |\Psi\rangle = \langle \prod_{p'} Q_{p'} \prod_p H_p V_p \rangle_{\text{virt}} |\text{phys}\rangle,
\end{equation}
where $\ket{\text{phys}}$ $\left(\ket{\text{virt}}\right)$ is the vacuum of the physical (virtual) space. The method used in \ref{sec:rev} can now be applied directly to complete the TN representation of the decohered state: simply modify $Q$ by introducing an ancilla to $T$ that contracts on-site with the bra layer:
\begin{equation} \label{eq:fermion T}
    T^k_{\alpha\beta\gamma\delta \omega} = \begin{cases} T^k_{\alpha\beta\gamma\delta} & \omega=0, \\ 0 & \omega=1. \end{cases}
\end{equation}
The index $\omega$ is associated with an ancillary complex fermion $w$, but this does not need to be included in the modification of $Q$ since the only terms that contribute are those with $\omega=0$, i.e. $w$ is effectively homogeneous in the $\Z_2$-graded Hilbert space. $h_{\omega\omega'}$ then connects $T$ in the ket layer to $T^*$ in the bra layer as before. defining $F=\langle \prod_{p'} Q_{p'} \prod_p H_p V_p \rangle_{\text{virt}}$, the TN representation of the decohered state is:
\begin{equation}
    \rho_f=\left( \prod_lh\right) F |\text{phys}\rangle \langle\text{phys}| F^\dagger,
\end{equation}
with $Q_p$ in $F$ containing the modified $T$ in (\ref{eq:fermion T}).

\section{Supplement to \texorpdfstring{$\mathbb{Z}_3$}{Z3} Numerics} \label{sec:Z3 numerics supplement}

\begin{figure}
\centering
    \begin{subfigure}[b]{0.95\columnwidth}
        \centering
        \includegraphics[width=\linewidth]{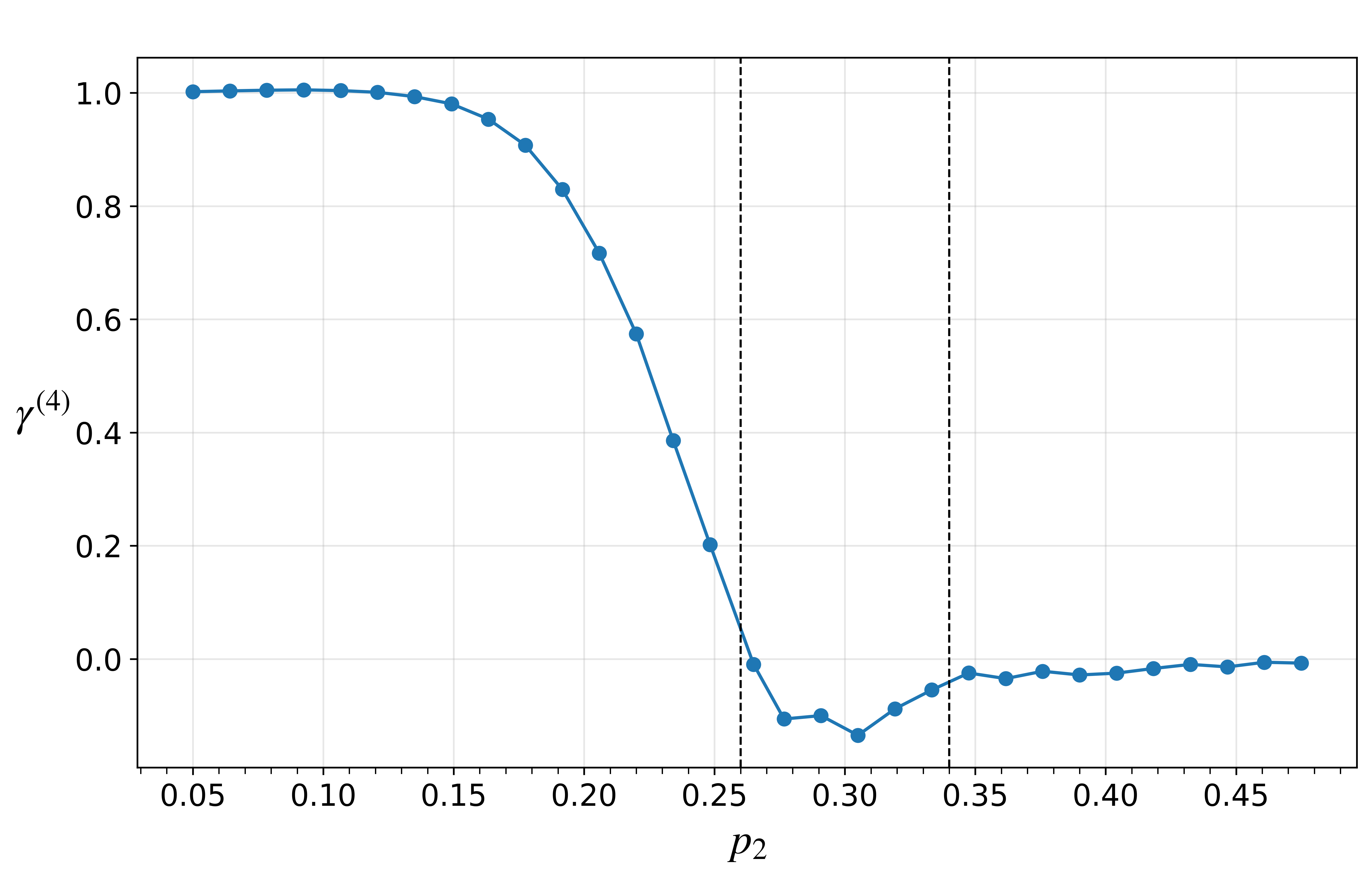}
        \caption{}
        \label{fig:TEN}
    \end{subfigure}
    \begin{subfigure}[b]{0.95\columnwidth}
        \centering
        \includegraphics[width=\linewidth]{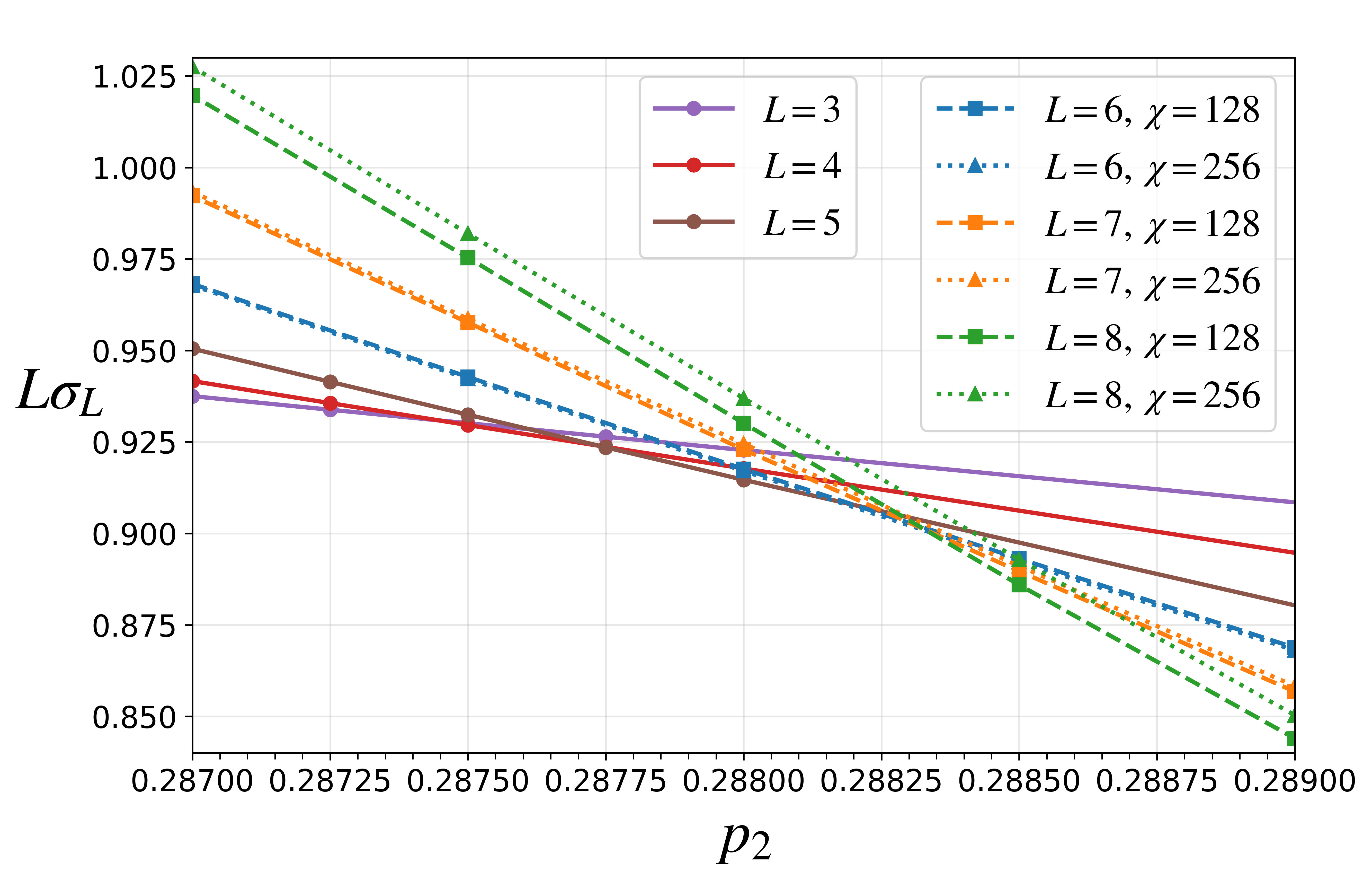}
        \caption{}
        \label{fig:LsigmaL zoom-in}
    \end{subfigure}
    \captionsetup{justification=raggedright, singlelinecheck=false}
    \caption{\footnotesize{(a) R\'enyi-4 TEN  (normalized by $\text{log}\;3$ for $\mathbb{Z}_3$) for $L=8,\;p_1=0.05$ as a function of $p_2$, with bond dimension truncation of $\chi=256$. The dashed lines indicate the bounds for expected transition point. (b) Zoom-in on the crossing region of Fig. \ref{fig:LsigmaL}. The crossings do not perfectly overlap due to finite size effects, as can be seen for $L \leq5$. The effect of bond dimension truncation is small for the purposes of the rough estimate.}}
\end{figure}

\begin{figure}
\centering
    \begin{subfigure}[b]{0.95\columnwidth}
        \centering
        \includegraphics[width=\linewidth]{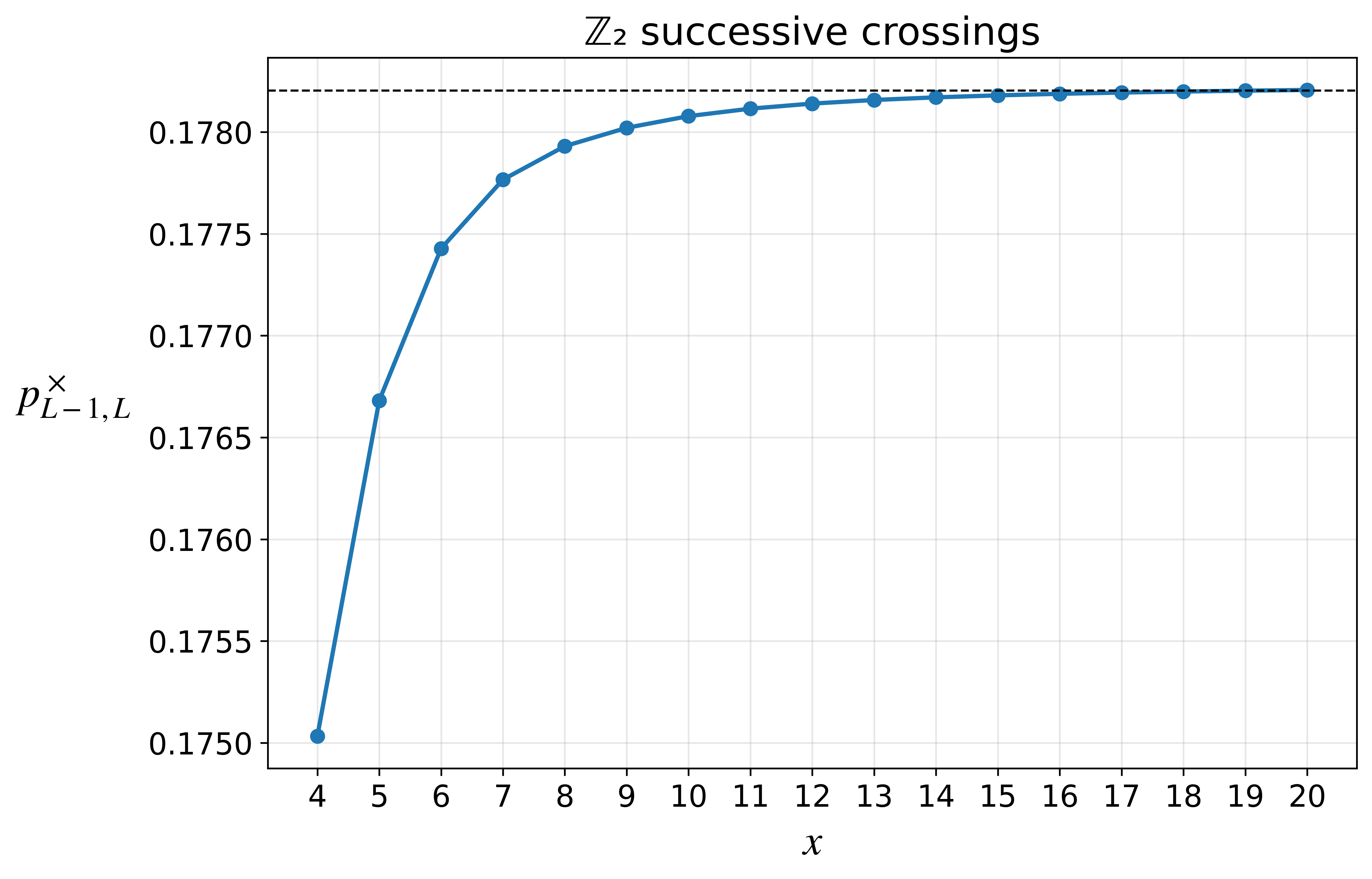}
        \caption{}
        \label{fig:Z2 crossings}
    \end{subfigure}
    \begin{subfigure}[b]{0.95\columnwidth}
        \centering
        \includegraphics[width=\linewidth]{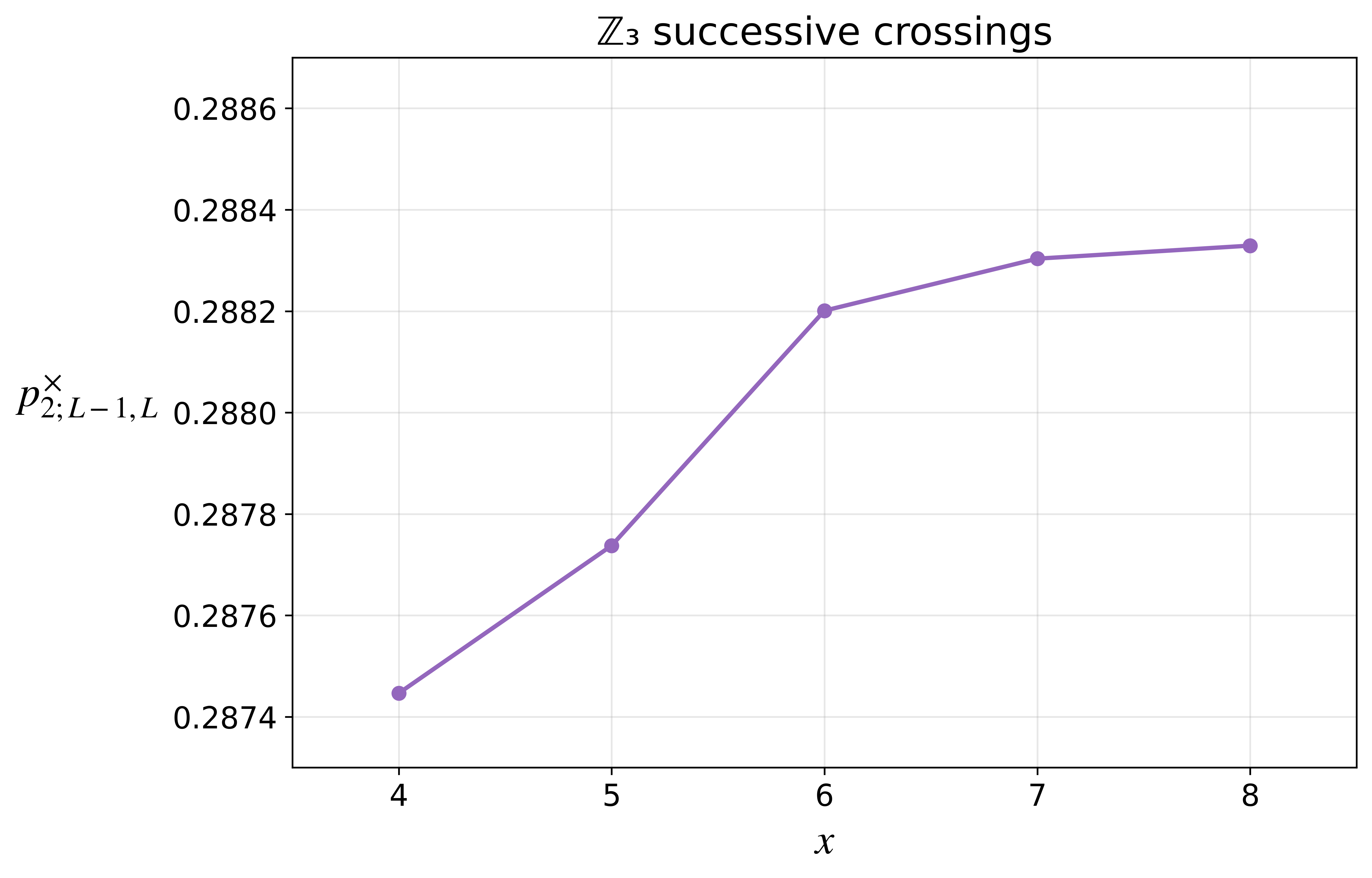}
        \caption{}
        \label{fig:Z3 crossings}
    \end{subfigure}
    \captionsetup{justification=raggedright, singlelinecheck=false}
    \caption{\footnotesize{(a) crossing $p$ values of $L\sigma_L$ (as defined in section \ref{sec:Z3 TEN and QRE}, with $n=2$) for successive system pairs of sizes $x$ in the $\mathbb{Z}_2$ toric code with $m$ decoherence. The crossings asymptotically approach the thermodynamic transition point as determined by KW duality. Power-law fitting with residual-sum minimization yields a predicted transition of $p_c\approx0.178220$, with exponent $\lambda\approx3.42$. (b) A similar plot for the $p_2$ crossings for the channel defined in section \ref{sec:Z3 TEN and QRE} for $\mathbb{Z}_3,n=4$. $L>5$ is truncated to bond dimension $\chi=128$. The predicted transition with the same method as before yields a predicted transition at $p_c\approx0.288$, with exponent $\lambda\approx8.50$.}}
\end{figure}

This appendix provides additional details to section \ref{sec:Z3 numerics}, starting with a comparitive calculation of the R\'enyi topological entanglement negativity (TEN), 
followed by more detailed diagnostics for determining the quantum relative entropy (QRE) transition.  

The TEN \cite{PhysRevA.65.032314,PhysRevA.88.042319} can serve as a diagnostic of the decoherence transition \cite{PRXQuantum.5.020343,cai2026entanglementnegativitydecoheredtopological,Mong_unpublished}, and is expected to jump from $\log 3$ to $0$ at the decoherence transition: below the error threshold, $\mathbb{Z}_3$ toric code contains $9$ types of Abelian anyons, and above the error threshold the post-decohered state host no topological quantum memory. More generally, the Rényi-$2n$ version of entanglement negativity with respect to subregion $A$ is defined as
\begin{equation}\label{eq:REN}
    \mathcal{E}^{(2n)}_A(\rho)=\frac{1}{2-2n}\log\frac{\text{Tr}(\rho^{T_A})^{2n}}{\text{Tr}\rho^{2n}},
\end{equation}
where $A$ is a subregion, and $\rho^{T_A}$ denotes the partial transpose of $\rho$ in $A$. The von Neumann entanglement negativity is obtained by analytically continuing the above to the $2n\rightarrow 1$ limit \cite{PhysRevLett.109.130502}. It's straightforward to see that the R\'enyi-2 entanglement negativity vanishes trivially, and therefore the first nontrivial index is $2n=4$. The goal of the TEN calculation is to provide a rough inital bound to the transition before a more detailed and computationally heavy analysis using QRE is performed. 

We use the given tensor network formalism in combination with transfer matrix methods, and extract the topological component $\gamma^{(4)}$ of the EN using a Kitaev-Preskill-like tripartition \cite{PhysRevLett.96.110404,PRXQuantum.5.020343}. Going off the symmetric line $p_1 \neq p_2$, we calculate for $L=8$ the TEN on the line $0.05\leq p_1\leq0.475,\;p_2=0.05$. The result is shown in Fig. \ref{fig:TEN}, which roughly bounds the transition to the region with pronounced finite size effects within $p_2\in[0.26,0.34]$.  

We now present in more detail the method used for determining the QRE transition within this bounded region.  In Fig. \ref{fig:LsigmaL zoom-in}, we zoom in on the mutual crossing area seen in Fig. \ref{fig:LsigmaL} and show the effects of varying bond dimension for $L>5$. The transition point can be determined by fitting the crossings of successive system sizes to a functional form $p_L=p_c+A L^{-\lambda}$ consistent with an expected continuous transition. Let $x$ be the larger of a given pair, e.g. $x=4$ is the crossing of $L=3,4$. Fig. \ref{fig:Z2 crossings} shows crossing curves as a function of $x$ for the $\mathbb{Z}_2,\;n=2$ case for demonstrative purposes; the fit minimizes the sum of squared residuals $\sum_j (p_j-p_{L_j})^2$, resulting in a predicted transition $p_c\approx0.178208$, which only deviates from the analytical transition known from Kramers-Wannier (KW) duality $p_\text{KW}=\frac{1}{2}\left( 1-\sqrt{\sqrt{2}-1} \right)\approx0.178203$ \cite{PhysRev.60.252,bao2023mixedstatetopologicalordererrorfield} on the order of $10^{-5}$. Fig. \ref{fig:Z3 crossings} shows the same method applied to the particle-hole asymmetric channel in $\mathbb{Z}_3$ toric code at $n=4$, where $\chi=128$ for $L>5$. The same fitting procedure results in a predicted transition at $p_c\approx0.288$. Naturally, the predicted location of the transition point is dependent on the choice of bond dimension truncation, though calculations with $\chi=256$ and $\chi=500$ show shifts in the predicted point on the order of $10^{-4}$, similar in size to the shift away from the KW-determined transition due to finite-size effects in the $\mathbb{Z}_2$ case.

\end{appendix}

\bibliography{references}
\end{document}